\documentclass[
notitlepage,             
prr,                     
superscriptaddress,      
nofootinbib,             
twocolumn,              %
floatfix]                
{revtex4-2}              

\usepackage{amssymb}
\usepackage{graphicx}
\usepackage{amsmath}
\usepackage{hyperref}
\usepackage{amsfonts}
\usepackage{natbib}
\usepackage{lipsum}
\usepackage{tabularx}
\usepackage{xcolor}
\usepackage{oubraces}
\usepackage{dsfont}
\usepackage{titlesec}
\usepackage{transparent}
\usepackage{tocloft}
\usepackage{color}
\usepackage{accents}
\usepackage{soul}

\DeclareRobustCommand{\rchi}{{\mathpalette\irchi\relax}}
\newcommand{\irchi}[2]{\raisebox{\depth}{$#1\chi$}} 

\begin{document}
\title{Quantum Rayleigh problem and thermocoherent Onsager relations}

\author{Onur Pusuluk}
\affiliation{Department of Physics, Ko\c{c} University, Sar{\i}yer,
\.{I}stanbul, 34450 Turkey}
\author{\"{O}zg\"{u}r E. M\"{u}stecapl{\i}o\u{g}lu}
\affiliation{Department of Physics, Ko\c{c} University, Sar{\i}yer,
\.{I}stanbul, 34450 Turkey}

\begin{abstract}
The role of quantum coherence and correlations in heat flow and equilibration is investigated by exploring the Rayleigh's dynamical problem to equilibration in the quantum regime and following Onsager's approach to thermoelectricity. Specifically, we consider a qubit bombarded by two-qubit projectiles from a side. For arbitrary collision times and initial states, we develop the master equation for sequential and collective collisions. By deriving the Fokker-Planck equation out of the master equation, we identify the quantum version of the Rayleigh's heat conduction equation. We find that quantum discord and entanglement shared between the projectiles can contribute to genuine heat flow only when they are associated with so-called heat-exchange coherences. Analogous to Onsager's use of Rayleigh's principle of least dissipation of energy, we use the entropy production rate to identify the coherence current. Both coherence and heat flows can be written in the form of quantum Onsager relations, from which we predict coherent Peltier and coherent Seebeck effects. The effects can be optimized by the collision times and collectivity.
Finally, we discuss some of the possible experimental realizations and technological applications of the thermocoherent phenomena in different platforms.
\end{abstract}

\maketitle

\section{INTRODUCTION}

\begin{figure}[b] \centering
        \includegraphics[width=.47\textwidth]{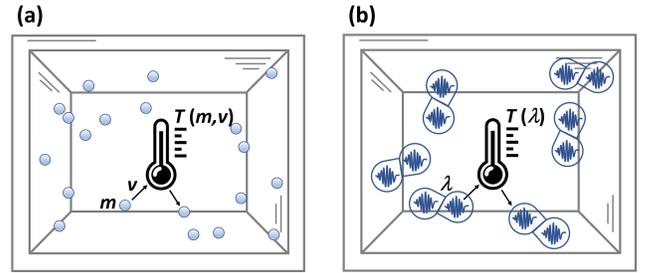}
        \caption{Three-dimensional illustration of (a) Rayleigh's one-dimensional collision model~\cite{rayleigh1891} in a box filled with a gas at some temperature and (b) its extension into the quantum regime where particles also behave as waves. The value $T$ shown in the thermometer depends on the classical degrees of freedom (DOF) of individual projectiles, mass $m$ and velocity $v$ in the former. Quantum projectiles consist of a pair of locally thermal two-level systems whose classical DOF are the population of the energy levels. These projectiles have an additional DOF, i.e., a quantum DOF represented by $\lambda$, which can also affect the rate of heat exchange with the thermometer.}
        \label{Fig_Rayleigh}
\end{figure}

In his 1891 paper~\cite{rayleigh1891}, Lord Rayleigh focused, for the first time, on the whole dynamical process by which a thermodynamic steady state is attained in a gas of particles. He captured the essential physics by developing a simplified model where tiny and light projectiles collide with heavy masses at rest (Fig.~\ref{Fig_Rayleigh} (a)). A further limitation of his studies on the one-dimensional bombardment of the masses by projectiles allowed him to reveal the role of the environment's initial conditions in the heat conduction. In particular, the mass and velocity of the projectiles striking the system appeared explicitly in the thermal diffusivity coefficient of Rayleigh's heat equation. This equation forms the basis for Onsager reciprocal relations between heat and charge flows~\cite{1931_Onsager}, which play the central role in thermoelectricity~\cite{2015_ThElec}.

This paper builds on the pioneering works of Lord Rayleigh and Onsager. We focus on the role of the initial quantum properties of the environment in heat conduction (Fig.~\ref{Fig_Rayleigh} (b)). Specifically, we simplify the complex problem of heat conduction dynamics in a gas of coherent quantum particles to a one-dimensional collision model~\cite{2002_PRL_88_097905, 2014_PRA_90_032111_CB, CM_2015_PRA_Ciccarello, CM_2017_QMeasQMetrol_Ciccarello_CMsInQOptics, 2017_PRA_96_032111_CB, CM_2018_PRA_EntangledBaths, 2020_arXiv_2006_12848}. Such a simplification allows us to derive the heat conduction equation with the explicit distinction of quantum coherent and purely thermal contributions, which turns out to be additive. Accordingly, we are able to establish analogs of Onsager relations between heat flow $\mathcal{J}_h$ and quantum coherence flow $\mathcal{J}_c$, which read
\begin{eqnarray}\label{eq:thermocohOnsgager}
\mathcal{J}_h&=&L_{hh}\,\Delta (1/T) + L_{hc}\,\Delta (-C)/T,\\
\mathcal{J}_c&=&L_{ch}\,\Delta (1/T) + L_{cc}\,\Delta (-C)/T ,
\end{eqnarray}
where $L_{hh}$ and $L_{cc}$ are the heat and quantum coherence conductivities, under temperature and quantum coherence differences $\Delta T$ and $\Delta C$, respectively. The interference of heat and coherence flows is reflected by the emergence of $L_{hc}$ and $L_{ch}$ coefficients that we dub as thermocoherent coefficients. We rigorously show that $\mathcal{J}_c$ delivered by the quantum projectiles is completely received as heat by the mass so that $\mathcal{J}_c = \mathcal{J}_h$ and reciprocity $L_{hc}=L_{ch}$ holds. Thermocoherent Onsager equations immediately reveal the possibility of mutual induction or control of heat and coherence flows. For example, heat flow between equal temperature reservoirs by a quantum coherence potential gradient, as an extension of the Peltier effect, or coherent current generation by thermal gradient as an extension of the Seebeck effect, are possible thermocoherent phenomena that can be predicted.
Remarkably, further phenomenological particle current terms could be introduced in the thermocoherent relations.

The systematic investigation of thermocoherent coefficients and quantum coherence conductivity that we present here can be significant for the  growing body of research on the thermodynamic means of generating and protecting quantum coherence and correlations
(QCCs)~\cite{2015_NewJPhys_Brunner_T2Ent, 2015_NJP_Brunner_TDCostOfCorrelations, 2017_PRA_Ozgur, 2018_QIP_EntIn2QubitsWith2CommonBaths, 2018_Quantum_BrunnerT2Ent, 2019_PRE_T2QCoh, 2020_PRA_Brunner_AutonomousMultipartiteEntEngines}, as well as for the studies of QCCs to manipulate thermodynamic energy transfer~\cite{AHF_2008_PRE_Partovi, Lutz2009, AHF_2010_PRE_JenningsAndRudolph, 2016_Entropy_Ozgur, 2018_PRE_RoleOfQCohInHT, AHF_2019_QuantumSciTechnol_Petruccione_AppTs, AutoThermalMach_2019, 2019_npj, 2019_PRE_Ozgur_Multiatom, 2019_PRL_Esposito_TDofInfoFlow, 2019_PRA_CohNeeded4HF, AHF_2019_NatCommun_Lutz, 2019_OSID, AHF_2019_PhysRevResearch_Petruccione, 2019_PRL_Esposito_PiWithBathCoh,  2020_PRA_HorizantolCohAndPops, AHF_2020_arXiv_IonTraps, 2020_arXiv_2006_01166}.

Another fundamental question that we ask here is which QCCs take a role in thermocoherent effects and heat conduction in a quantum gas of particles.  To address this question, we consider all particles as two-level systems (qubits) in our collision model, and take the projectiles as pairs
of qubits in generic quantum and classical correlated states. After introducing the quantum Rayleigh problem, describing the collisions between the projectiles and the target qubit in Sec.~\ref{Sec_qRayleighCM}, we will present the the projectile states explicitly in Sec.~\ref{Sec_ProjectileStates}. Section~\ref{Sec_Thermalization} develops the thermalization dynamics of the target qubit in terms of the master equation (Sec.~\ref{Sec_MasterEq}), which is used to determine the quantum heat current (Sec.~\ref{Sec_FPform}). The quantum coherence current is identified in Sec.~\ref{Sec_CohCurrent} and quantum thermocoherent Onsager relations are established in Sec.~\ref{Sec_Onsager}. We discuss the experimental and application significance of the results in Sec.~\ref{Sec_Apps} and conclude in Sec.~\ref{Sec_Conclusions}.
\section{QUANTUM RAYLEIGH PROBLEM}\label{Sec_qRayleighCM}

\begin{figure}[t] \centering
	\includegraphics[width=.45\textwidth]{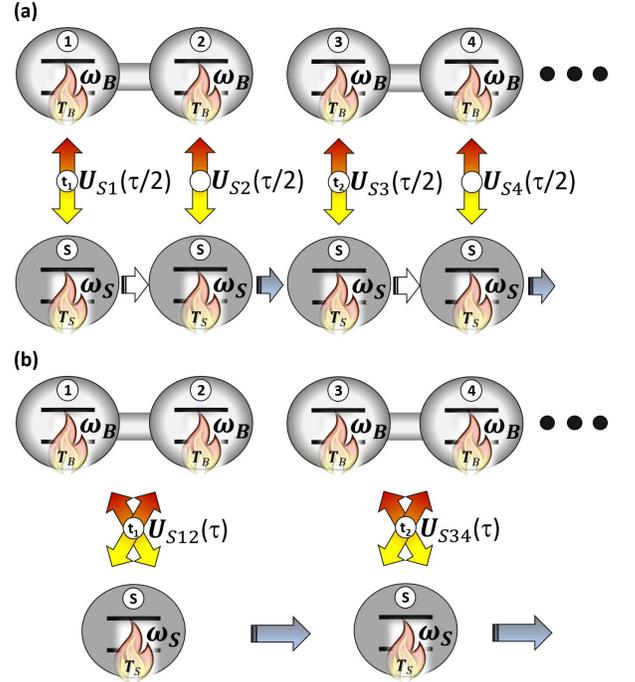}
	\caption{Quantum version of Rayleigh's collision model with initially correlated bath qubits labeled $\{1, 2, 3,...\}$. Identical and non-interacting bath qubits are locally at thermal equilibrium at temperature $T_B$, while sharing either classical correlation, quantum discord, or quantum entanglement in pairs (linked circles). Projectiles consisting of such pairs strike the target qubit $S$ at random times $t_j$, which has an energy level separation of $\omega_S$ and an initial temperature $T_S$. $S$ interacts with each bath qubit in the $j$th pair either (a) sequentially or (b) collectively. The whole collision with a single projectile takes over a time interval of duration $\tau \ll t_{j+1} - t_j$. Sequential collision is described by the product of two-qubit operators $U_{SB_{2j}}(\tau/2)U_{SB_{2j-1}}(\tau/2)$,while collective collision is described by a single three-qubit operator $U_{SB_{2j-1}B_{2j}}(\tau)$.}
	\label{Fig_Model_SingleBath}
\end{figure}

When Lord Rayleigh examined the collisional route to equilibration,
as depicted in Fig~\ref{Fig_Rayleigh} (a), in terms of a simplified
one-dimensional bombardment of a massive atom at rest by tiny projectiles,
he stated some corollaries, which are natural for a classical physicist. In particular
he wrote that ``If we suppose that the projectiles are dispatched in pairs of
closely following components, we should expect that the effect would be
the same as of a doubling of the mass"~\cite{rayleigh1891}. Here,
we explicitly generalize the Rayleigh problem to the case of pairwise
projectiles, which will not be a trivial case of doubling of mass, if the projectiles have a wavelike nature, in particular, carry a pairwise quantum
coherence and correlations, as illustrated in Fig~\ref{Fig_Rayleigh} (b).
For brevity of terminology, we call the target qubit, representing
the stationary massive (thermometer) atom, only as qubit and the pair of
qubits bombarding it as projectiles.

The projectiles (marked with $B$) are identical copies each prepared in an arbitrary, possibly nonthermal state, while the qubit (marked with $S$) is initially in a thermal equilibrium state
\begin{equation}
	\rho_S(0) = q_g |g \rangle \langle g | + q_e |e \rangle \langle e |,
\end{equation}
where $q_{g/e}=\exp(- \beta_S E_{g/e}) / [\exp(- \beta_S E_g) + \exp(- \beta_S E_e)]$ and $\beta_S = 1/k_\mathrm{B} T_S$ with $k_\mathrm{B}$ being the Boltzmann constant.
The qubit Hamiltonian reads
\begin{equation}\label{eq:qubitH}
	H_S = E_g |g \rangle \langle g | + E_e |e \rangle \langle e |.
\end{equation}

Projectiles arrive at the qubit at times $t_j$ and the collision
time is fixed to $\tau$. We envision two different bombardment scenarios.
In the first case, the qubit interacts with each projectile qubit
sequentially, as illustrated in Fig.~\ref{Fig_Model_SingleBath} (a).
The collision at a time $t_j$ is then described by
\begin{equation}\label{Eq_Us}
	U(\tau) = U_{SB_{2j-1}}(\tau/2) \, U_{SB_{2j}}(\tau/2),
\end{equation}
where each unitary $U_{SB_{j}}$ is generated by an interaction Hamiltonian
of the form
\begin{eqnarray}\label{Eq_Hsb}
		H_{SB_j} =
		J ( \sigma_x^{(S)} \sigma_x^{(B_j)} + \sigma_y^{(S)} \sigma_y^{(B_j)} ).
\end{eqnarray}
Here, $J$ is the coupling constant, and $\sigma_x,\sigma_y$ are the Pauli matrices.

The second scenario (Fig.~\ref{Fig_Model_SingleBath}(b)) focuses on a collective collision described by the following evolution
\begin{equation}\label{Eq_Uc}
	U(\tau) = U_{SB_{2j-1}B_{2j}}(\tau) = e^{- \frac{i \tau}{\hbar} (H_{SB_{2j-1}}+H_{SB_{2j}})} ,
\end{equation}
where the qubit collides with the both qubits in the $j$th projectile simultaneously. The process is described by the Hamiltonian in Eq.~(\ref{Eq_Hsb}).
The total energy is preserved in both collision scenarios, yet the energy
exchange between the projectiles and the qubit varies. Accordingly,
both the equilibration and heat conduction are sensitive to the collective
nature of collisions.
\section{INITIAL STATES OF THE PROJECTILES}\label{Sec_ProjectileStates}

\begin{figure}[t] \centering
        \includegraphics[width=.45\textwidth]{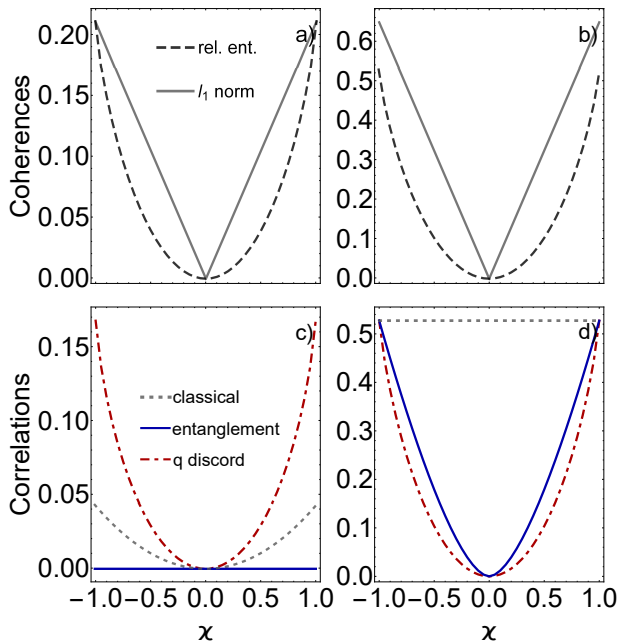}
        \caption{Amount of the quantum coherences and correlations found in quantum-correlated local thermal bath states $\rho_D$ (left panels) and $\rho_E$ (right panels) with ${\footnotesize{\rchi}} = \lambda/\lambda_{max} = \mu/\mu_{max}$. $E_g = 1$, $E_e = 2$, and $\beta_B = 2$ in all four panels. In the upper panels, the relative entropy of coherence~\cite{Plenio-2014} and $l_1$ norm of coherence~\cite{Plenio-2014} are shown, respectively, by dashed black and gray curves. Dotted gray, dash-dotted red, and blue curves in the lower panels correspond to classical correlations~\cite{Vedral-2001}, quantum discord~\cite{Vedral-2001, Zurek-2002}, and entanglement of formation~\cite{Wooters-1996}.}
        \label{Fig_QBath}
\end{figure}

Projectiles are assumed to be prepared in a thermal equilibrium state
at temperature $T_B$ first; then they are either correlated classically or
injected with some quantum coherence. Depending on which states carry the quantum
coherence, the projectile can have profoundly distinct quantum correlations.
Specifically, we will examine three different states of the projectiles before
the collision with the qubit, which are expressed as
\begin{subequations}\label{Eq_rho_B}
\begin{align}
\rho_C &= p_g |gg \rangle \langle gg | + p_e |ee \rangle \langle ee | , \label{Eq_rho_2CC}\\
\rho_D &= (p_g |g \rangle \langle g | + p_e |e \rangle \langle e |)^{\otimes 2} + (\lambda |ge\rangle \langle eg| + \mathrm{h.c.}) , \label{Eq_rho_2QD} \\
\rho_E &= p_g |gg \rangle \langle gg | + p_e |ee \rangle \langle ee | + (\mu |gg\rangle \langle ee| + \mathrm{h.c.}) , \label{Eq_rho_2QE}
\end{align}
\end{subequations}
where $|\lambda|\le \lambda_{\text{max}}:=p_g p_e$ and $|\mu| \leq \mu_{\text{max}}:=\sqrt{p_g p_e}$ because of the positivity of $\rho_D$ and $\rho_E$, respectively, and $p_{g/e}=\exp(-\beta_B E_{g/e})/[\exp(-\beta_B E_g) +\exp(-\beta_B E_e)]$ with $\beta_B=1/k_\mathrm{B} \, T_B$. Projectile qubits are identical with the qubit so that their Hamiltonian $H_B$ is the same as in Eq.~(\ref{eq:qubitH}). Injected coherence in neither of these states destroys the local thermal equilibrium of each qubit of the projectile. Their reduced (local) states are given by the Gibbs state $\exp[- \beta_B H_B]/\mathrm{tr}[\exp[- \beta_B H_B]]$. Accordingly, we can associate $T_B$ as a local temperature to each projectile qubit. This is the first criterion behind the choice of the projectile states $\rho_C$, $\rho_D$, and $\rho_E$ in Eq.~(\ref{Eq_rho_B}). Second, each of these three state families illustrates a different kind of correlation as discussed below.

Projectile pairs that share only classical correlations are described by $\rho_C$~\cite{Vedral-2001}. Pairs in $\rho_D$ and $\rho_E$ possess nonclassical correlations that can be well captured by so-called quantum discord~\cite{Vedral-2001, Zurek-2002}, though they have quite different natures. $\rho_D$ is a separable state like $\rho_C$ but its preparation using incoherent operations requires the consumption of a local quantum coherence initially found in the projectiles~\cite{Coh2Discord}. It describes a sensitivity to local dynamics in such a way that local measurements disturb the pre-measurement state of both projectiles. Unlike the classical correlations contained in $\rho_C$, the total correlations carried in this state cannot be extracted by any local quantum measurements~\cite{Vedral-2001, Zurek-2002, QRT_2019}. On the other hand, $\rho_E$ is an entangled state, i.e. it cannot be factorized into a product state. Its quantum correlations can be measured not only by the quantum discord but also by the entanglement of formation~\cite{Wooters-1996}, which always gives zero for $\rho_D$. To emphasize this difference, whenever we use the term ``discordant projectile state'' in what follows, we will refer to $\rho_D$. On the other hand, ``entangled projectile state'' will stand for $\rho_E$.

Quantum coherence and correlations in these states are compared in Fig.~\ref{Fig_QBath}. The nonzero values of $\lambda$ and $\mu$ generate quantum coherence and discord respectively in $\rho_D$ and $\rho_E$, which differ only quantitatively (cf.~Fig.~\ref{Fig_QBath} (a)-(d)). Quantum correlation in $\rho_D$  becomes a maximum in the case of $|\lambda|$ is equal to $\lambda_{\mathrm{max}}$ and decreases when $\lambda$ approaches zero (cf.~Fig.~\ref{Fig_QBath} (c)). In the limit of vanishing $\lambda$, the pair becomes uncorrelated. Similarly, entangled state $\rho_E$ reduces to classically correlated state $\rho_C$ at $\mu = 0$. Simple monotonic behavior of coherence and correlations (with either positive or negative parameters) and lack of entanglement in the discordant state allows us to investigate influence of QCCs in thermocoherent relations explicitly. This is the last criterion that determines the three projectile states given in Eq.~(\ref{Eq_rho_B}).

\begin{table*}
\centering \caption{The coefficients in the time-dependent qubit state~(\ref{Eq_rho_S}) are defined depending on the collision scenario and correlation type under consideration. Note that the value of $\alpha$ given for sequential collisions converges to unity in the weak coupling or short interaction limit. To avoid potential confusion, it should also be emphasized that $\alpha$ is a function of collectivity, strength, and duration of the collisions, but the dependence on these parameters is omitted in the text for the sake of readability.}
\begin{tabular}{p{3.5cm} p{3.5cm} p{2cm} p{4cm} p{4cm}{c}}
\hline
Scenario & Correlations & $\alpha$ & $\Gamma_{g/e}$ & $\gamma(t)$ \\
 \hline
Sequential Collision & Quantum Discord & $\frac{2 \cos(J \tau)}{1 + \cos^2(J \tau)}$ & $p_{g/e} \frac{1}{1 + 2 \, \alpha \lambda}+\frac{\alpha \lambda}{1 + 2 \, \alpha \lambda}$ & $e^{-\frac{p t}{\alpha}(1 + 2 \, \alpha \lambda)\sin(J \tau)\sin(2 J \tau)}$
\\
 & Others & $\frac{2 \cos(J \tau)}{1 + \cos^2(J \tau)}$ &  $p_{g/e}$ & $e^{-\frac{p t}{\alpha}\sin(J \tau)\sin(2 J \tau)}$ \\
 Collective Collision & Quantum Discord & 1 & $p_{g/e} \frac{1}{1 + 2 \, \alpha \lambda}+\frac{\alpha \lambda}{1 + 2 \, \alpha \lambda}$ & $e^{- \frac{p t}{\alpha} \, (1 + 2 \, \alpha \lambda) \sin^2(2 \sqrt{2} J \tau)}$ \\
  & Others & 1 & $p_{g/e}$ & $e^{- \frac{p t}{\alpha} \sin^2(2 \sqrt{2} J \tau)}$ \\
\hline
\end{tabular}
\label{Table_rho}
\end{table*}

\section{DYNAMICS TO STEADY STATE} \label{Sec_Thermalization}
\subsection{Master Equation and Anomalous Heat Current} \label{Sec_MasterEq}

We derive a master equation~(\ref{Eq_meq}) for arbitrary collision times and initial states in Sec.~\ref{Appendix::MEq1} in the Appendix. It is straightforward to solve this equation analytically for both sequential and collective collision scenarios. In the Schr\"{o}dinger picture, the solution takes the following general form
\begin{eqnarray}\begin{aligned}\label{Eq_rho_S}
\rho_S(t) &= \big(\Gamma_g (1 - q_e \gamma(t)) + \Gamma_e q_g \gamma(t)\big) | g \rangle \langle g | \\
&\qquad + \big(\Gamma_e (1 - q_g \gamma(t)) + \Gamma_g q_e \gamma(t)\big) | e \rangle \langle e | ,
\end{aligned}\end{eqnarray}
where $\Gamma_{g/e}$ and $\gamma(t)$ are defined in Table~\ref{Table_rho}, depending on the collision scenario and the correlation type under consideration.  In particular, classical correlations and quantum entanglement in the projectile are not found to play any role in the target qubit's open system dynamics. Recall that the discordant pair becomes uncorrelated when $\lambda$ goes to zero, and the coefficients in Table~\ref{Table_rho} turn out to be independent of the correlation type in this limit.

The steady state of Eq.~(\ref{Eq_meq}) is
\begin{eqnarray}\begin{aligned}\label{Eq_rho_Sinf}
\rho_S^\infty = \Gamma_g | g \rangle \langle g | + \Gamma_e | e \rangle \langle e | ,
\end{aligned}\end{eqnarray}
which is equal to the local (reduced) states of the qubits of the projectile in $\rho_C$
or in $\rho_E$ states. Neither classical correlations nor entanglement can make any difference in the thermalization of the qubit to the local temperature $T_B$. In Sec.~\ref{Sec_CohCurrent}, we will revisit the case of entangled projectiles. We will particularly show that these projectiles can affect the dynamics of the qubit and its steady state if the collision block is extended to include two projectile pairs instead of one.

\begin{figure}[b] \centering
        \includegraphics[width=.95\linewidth]{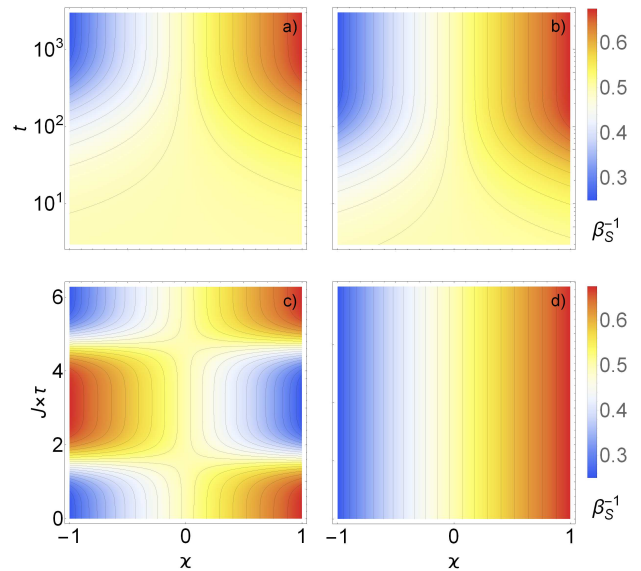}
        \caption{Temperature of the qubit bombarded with projectiles in discordant states $\rho_D$. Upper panels show $\beta_S^{-1}$ as a function of time $t$ and ${\footnotesize{\rchi}} = \lambda/\lambda_{max}$ in the weak-coupling limit ($J = 0.05/\tau$). Here, we work with dimensionless time with a unit system where $p = 1$. The value of $\beta_S^{-1}$ is explored for different coupling strengths when $t\rightarrow\infty$ in the lower panels. $E_g = 1$, $E_e = 2$, $\beta^{-1}_B = \beta^{-1}_S(t = 0) = 0.5$ in all four panels. The sequential (collective) collision model is considered in the left (right) column.}\label{Fig_Temps}
\end{figure}

On the other hand, Eq.~(\ref{Eq_rho_Sinf}) corresponds to a thermal state for collisions with discordant projectiles when $\lambda > - 1/ (2 \alpha)$ (cf.~ Table~\ref{Table_rho}). The qubit reaches a final temperature
\begin{equation}\label{Eq_T_Sinf}
T_{S}(\infty) = \frac{(E_g - E_e)/k_\mathrm{B}}{\ln[(p_e + \alpha \lambda)/(p_g + \alpha \lambda)]},
\end{equation}
which shows that the discordant projectiles thermalize the qubit to a temperature that is different than their local temperature $T_B$. $T_{S}(\infty)$ can be greater or less than $T_B$ depending on the sign of $\alpha \lambda$, as shown in Figs.~\ref{Fig_Temps} (c) and (d). In the high-temperature limit, we get $T_{S}(\infty) \approx T_B (1 + 2 \, \alpha \lambda)$. Note that $- 1/ (2 \alpha) < \lambda \leq p_g p_e$, $-1 \leq \alpha \leq 1$, and $p_g p_e \approx 1/4$ for high temperatures. The heating and cooling with
quantum discord depends on the sign of $\lambda$ for both collective and sequential collision scenarios. In the case of sequential interactions, however, the sign of
$\alpha(J\tau)$ also becomes a factor.

$T_{S}(\infty) \neq T_B$ allows for enhancement or reduction in the heat flow between the qubit and discordant bath. When $T_S(0) = T_{S}(\infty)$,  the qubit remains at its initial temperature so that the heat flow is inhibited, despite the temperature difference of $~2 \, \alpha \lambda \, T_B$.
Conversely, an anomalous heat emerges when $T_S(0) = T_B$.
The absence of an initial temperature gradient does not prevent the change in the temperature of the qubit given by
\begin{equation}\label{Eq_T_S}
T_{S}(t) = \frac{(E_g - E_e)/k_\mathrm{B}}{\ln\Big[\frac{p_e + \alpha \lambda (1 + \gamma(t)(p_e - p_g))}{p_g + \alpha \lambda (1 + \gamma(t)(p_g - p_e))}\Big]}.
\end{equation}
It goes to $T_B (1 + 2 \, \alpha \lambda)/(1 + 2 \, \alpha \lambda \gamma(t))$ in the high-$T_B$ limit. $T_S(t)$ is further reduced to $T_B (1 + 2 \, \alpha \lambda (1 - \exp[- p\, t \eta/\alpha]))$ for small values of $\lambda$. Here and in the following, $\eta$ stands for $\sin(J \tau) \sin(2 J \tau)$ and $\sin^2(2\sqrt{2} J \tau)$, respectively, for sequential and collective collisions. For the sake of readability, the dependence of $\eta$ on the collectivity, strength, and duration of the collisions is omitted in the text.

Thermalization of the target qubit to $T_{S}(\infty) \neq T_B$ does not
mean that there will be a persistent heat current between the projectile bath and the qubit; as we will explicitly calculate, the heat flow vanishes at the steady state. Nevertheless, the possibility of heating or cooling a qubit, in a gas of qubits initially at thermal equilibrium, by injecting pairwise discord to the rest of the qubits can be regarded as a transient thermocoherent effect.

\begin{figure}[t] \centering
        \includegraphics[width=.95\linewidth]{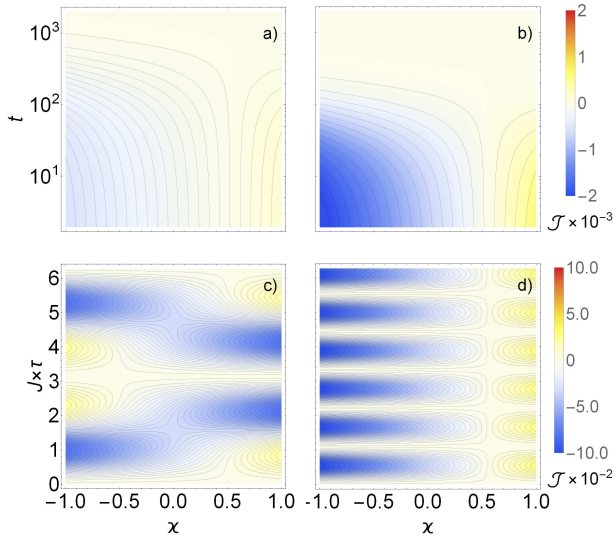}
        \caption{Net heat current between the qubit and discordant heat bath when $\beta_S^{-1}(0) = 0.6$ and $\beta_B^{-1} = 0.5$. The sequential (collective) collision model is considered in the left (right) column. Upper panels show $\mathcal{J}$ in the weak-coupling limit ($J = 0.05/\tau$), while its value for different coupling strengths is explored at $t = 0.1$ in the lower panels. The remaining parameters are the same as in Fig.~\ref{Fig_Temps}.}\label{Fig_CurrentsNet1}
\end{figure}

\begin{figure}[t] \centering
        \includegraphics[width=.95\linewidth]{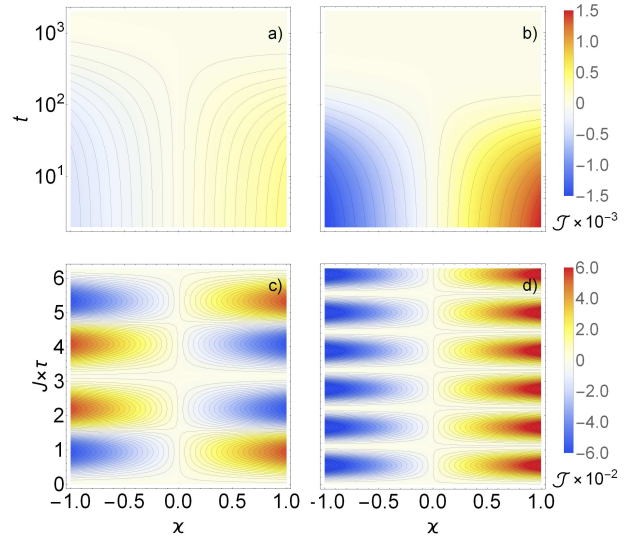}
        \caption{Anomalous heat current between the qubit and discordant heat bath when $\beta_S^{-1}(0) = \beta_B^{-1} = 0.5$. The sequential (collective) collision model is considered in the left (right) column. The upper panels show $\mathcal{J}$ in the weak-coupling limit ($J = 0.05/\tau$), while its value for different coupling strengths is explored at $t = 0.1$ in the lower panels. The remaining parameters are the same as in Fig.~\ref{Fig_Temps}.}\label{Fig_Currents1}
\end{figure}

\subsection{Quantum Heat Conduction} \label{Sec_FPform}

To investigate the heat flow, we use the quantum thermodynamic
definition~\cite{HeatCurrent},
\begin{eqnarray} \label{Eq_HeatCurrent}
\mathcal{J}(t) = \mathrm{tr}[H_S\,\frac{\mathrm{d}}{\mathrm{d}t}\rho_S(t)] ,
\end{eqnarray}
where a positive current means that heat flows from the bath of projectiles to the qubit and \textit{vice versa} when $\mathcal{J} < 0$. This is simply a consequence of the fact that the collisions are energy-preserving, which implies that they have zero work cost and that any energy exchange is pure heat. We further justify it more rigorously in the Appendix by rewriting the master equation~(\ref{Eq_meq}) in the Lindblad form in Appendix~\ref{Appendix::MEq2}, which is used to determine the quantum heat conduction equation via the Fokker-Planck equation in Appendix~\ref{Appendix::FPEq}.

Let us express the heat current explicitly, by substituting Eq.~(\ref{Eq_meq2}) into Eq.~(\ref{Eq_HeatCurrent}),
\begin{equation}\label{Eq_Js}
\mathcal{J}(t) = \mathcal{J}_0 \, \gamma(t)  ,
\end{equation}
where the time-independent amplitude is given by
\begin{equation}\label{Eq_Js0}
\mathcal{J}_0 = \big((q_g p_e - q_e p_g) + \alpha \lambda (q_g - q_e)\big)(E_e - E_g)\, \eta /\alpha ,
\end{equation}
and $\gamma(t)$ depends on $\alpha$, $\lambda$, and $J \tau$ (see Table~\ref{Table_rho}).

The term preceding $\alpha \lambda$ in $\mathcal{J}_0$ drops when $T_S(0)$ is set to $T_B$, and a transient anomalous heat current can be identified to be
\begin{equation}\label{Eq_JsP}
\mathcal{J}(t) = \lambda \, \eta \, (q_g - q_e)(E_e - E_g) \, \gamma(t) .
\end{equation}
which can only emerge in the case of discordant projectiles. In the high temperature limit it reduces to
\begin{equation}\label{Eq_Js2}
\mathcal{J}_h(t) \approx \gamma(t) \big(\tilde{L}_{hh} \Delta \beta - \tilde{L}_{hc} \, \beta_S(0) \, \Delta C\big) ,
\end{equation}
where $\Delta \beta = (\beta_S(0) - \beta_B)$, $\Delta C = \alpha(0 - 2 \, \lambda)$, and
\begin{equation}\label{Eq_OnsagerCoefs}
\tilde{L}_{hh} = \tilde{L}_{hc} = (E_e - E_g)^2 \, \eta / (4 \alpha) .
\end{equation}

We conclude that in addition to the initial local temperature difference between the projectiles and the qubit, quantum coherence in a discordant projectile acts as an additional source, analogous to chemical potential, to the heat flow. While the thermal gradient yields the normal heat flow direction between hot and cold bodies, the discordant source can give rise to anomalous heat behavior. For example, it can lead to inhibition of heat flow when these two terms cancel each other if $T_S(0) = T_S(\infty) \approx T_B (1 + 2 \, \alpha \lambda)$. On the other hand, only the second term survives when $T_S(0) = T_B$, and a heat flow persists between two bodies at equal temperatures. These examples are thermocoherent effects in the transient regime, and heat flow vanishes in the steady state.

Quantitative investigations of Eqs.~(\ref{Eq_Js})~and~(\ref{Eq_JsP}), are presented in Figs.~\ref{Fig_CurrentsNet1}~and~\ref{Fig_Currents1}, respectively. The net current disappears in spite of the presence of temperature gradient for some nonzero values of $\alpha \lambda$ in Fig.~\ref{Fig_CurrentsNet1}; the same values of $\alpha \lambda$ sustain a current in the absence of temperature gradient in Fig.~\ref{Fig_Currents1}. On the other hand, a comparison of Figs.~\ref{Fig_Temps}~and~\ref{Fig_Currents1} reveals that what is responsible for the qubit's temperature change in the absence of initial temperature gradient is the anomalous heat current defined in Eq.~(\ref{Eq_JsP}).

\subsection{Quantum Coherence Current}
\label{Sec_CohCurrent}

Extending Lord Rayleigh's principle of the least dissipation of energy, Onsager showed in Ref.~\cite{1931_Onsager} that the rate of increase of the entropy plays the role of potential in irreversible processes. Here, we consider the quantum version of this extension~\cite{2019_npj, 2019_PRL_Esposito_TDofInfoFlow, 2019_OSID, 2019_PRL_Esposito_PiWithBathCoh, 2020_arXiv_2009_07668} that relates the standard thermodynamic definition of the entropy production to the quantum-information-theoretical quantities
\begin{eqnarray}\begin{aligned}\label{Eq_EntropyPro}
\Pi(t) &= - \frac{\mathrm{d}}{\mathrm{d}t} S[\rho_S(t)||\rho_S^\infty] \\
&= - \mathrm{tr}[\mathcal{D}_h(\rho_S) \ln(\rho_S(t))] + \mathrm{tr}[\mathcal{D}_h(\rho_S) \ln(\rho_S^\infty)] \; \\
&\equiv \dot{S}_{\mathrm{vN}}[\rho_S(t)] + \Phi(t) ,
\end{aligned}\end{eqnarray}
where $S_\mathrm{vN}[\rho]$ is the von Neumann entropy that is equal to $- \mathrm{tr}[\rho \, \ln \rho]$, and $S[\rho||\sigma]$ is the quantum relative entropy which measures the distinguishability of the two density matrices $\rho$ and $\sigma$. The dynamical version of the second law of thermodynamics~\cite{2013_Entropy_15_02100_Kosloff} is always satisfied during the heat-exchange process under consideration since the entropy production rate $\Pi$ obtained in this way never becomes negative,
\begin{equation}\label{Eq_EntropyPro1}
\Pi(t) = \mathcal{J}(t) \big(\beta_S(t) - \beta_S(\infty)\big) \geq 0 \, ,
\end{equation}
even in the absence of initial temperature difference. This equation also says that the generalized thermodynamic force that generates the net heat current~(\ref{Eq_Js}) is simply how far the system's temperature is from the steady-state temperature. On the other hand, both $\beta_S(t)$ and $\beta_S(\infty)$ depend on the value of $\Delta C = - 2 \, \alpha \lambda$. To split the entropy production rate~(\ref{Eq_EntropyPro1}) into a term that
only depends on the temperature difference, and one that is induced by coherence as in Eq.~(\ref{Eq_Js2}), we can expand it in the high-temperature limit,
\begin{equation}\label{Eq_EntropyPro2}
\Pi(t)\approx \gamma(t) \Big(\Delta \beta \, \mathcal{J}_h(t)  - \beta_S(\infty) \,
\Delta C \, \mathcal{J}_c(t)\Big) ,
\end{equation}
where $\mathcal{J}_h(t)$ given in Eq.~(\ref{Eq_Js2}) is already properly identified as heat flow by our Lindblad and Fokker-Planck analyses in the Appendix. After dropping the multiplication of $\mathcal{J}(t)$ and its conjugate displacement $\gamma(t) \Delta \beta$ in $\Pi(t)$, we can identify the coherence flow $\mathcal{J}_{c}(t)$ driven by $-\gamma(t) \beta_S(\infty) \,
\Delta C$ as
\begin{equation}\label{Eq_Jc}
-\mathcal{J}_c(t) \equiv \gamma(t) \big(\tilde{L}_{ch} \Delta \beta - \tilde{L}_{cc} \, \beta_S(0) \, \Delta C\big) ,
\end{equation}
with $\tilde{L}_{cc} = \tilde{L}_{ch} = \tilde{L}_{hc}$. The minus sign preceding $\mathcal{J}_c(t)$ means that the heat and coherence currents are in opposite directions.

\begin{figure}[t] \centering
        \includegraphics[width=.95\linewidth]{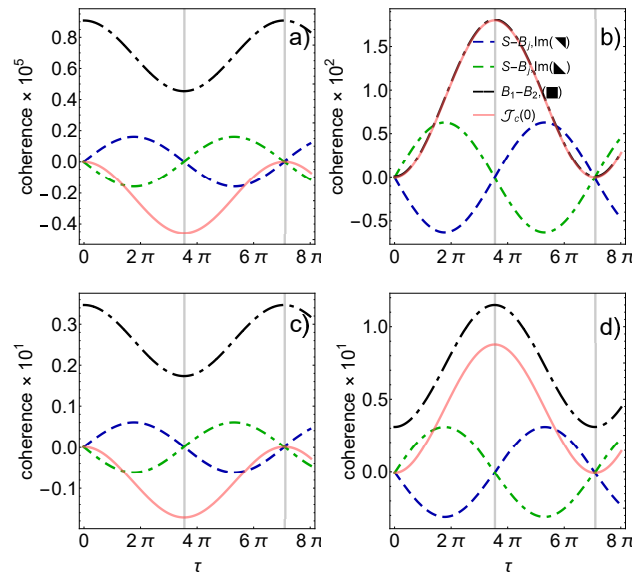}
        \caption{Macroscopic coherence current $\mathcal{J}_{c}(t=0)$ defined in Eq.~(\ref{Eq_Jc}) (pink solid curve) and microcurrents generated during the first collective collision of the qubit $S$ with a projectile pair. The change in the coherence shared between the projectile qubits $B_1$ and $B_2$ (double-dash-dotted black curve) exactly matches up with $\mathcal{J}_{c}(t=0)$. This coherence drops in $S$-$B_1$ and $S$-$B_2$ joint systems without a localization in any single-qubit subsystem and then returns back to the $B_1$-$B_2$ joint system. While all the elements of $\rho_{B_1B_2}$ are real, off-diagonal elements of $\rho_{SB_j}$ are purely imaginary. The dashed blue (dash-dotted green) curve shows the change in the imaginary part of the upper (lower) off-diagonal element of $\rho_{SB_j}$. $\lambda = \lambda_{max}$, $E_1 = 1$, $E_2 = 2$, $J = 0.05$ in all panels. The $\{\beta_B,\beta_S\}$ pair is (a) $\{10,10\}$, (b) $\{10,4\}$, (c) $\{4,4\}$, and (d) $\{4,2\}$. Note that $\mathcal{J}_{c}(t=0)$ is equal to twice the steady-state coherence current $\mathcal{J}_{c}$ given in Eq.~(\ref{Eq_JcSS}) when the second projectile bath shares the same temperature and coherence as the qubit.}\label{Fig_Jc}
\end{figure}

\begin{figure}[t] \centering
        \includegraphics[width=.95\linewidth]{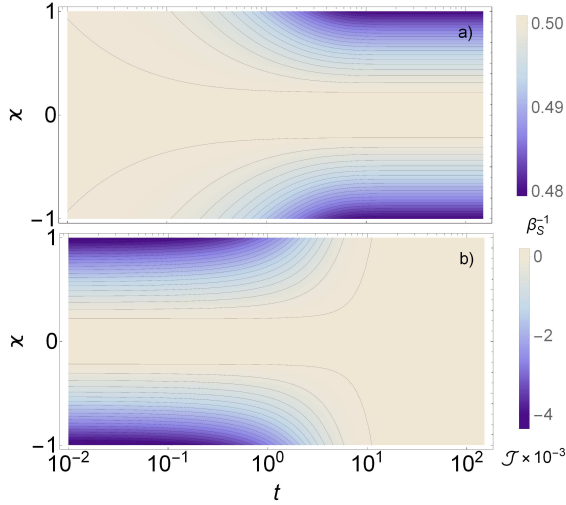}
        \caption{(a) Temperature of the qubit bombarded with projectiles in states $\rho_E \otimes \rho_E$ and (b) resulting anomalous heat current when $\beta_S^{-1}(0) = \beta_B^{-1} = 0.5$. $J = 0.2/\tau$, while the remaining parameters are the same as in Fig.~\ref{Fig_Temps}.}\label{Fig_4QE}
\end{figure}

As the qubit $S$ always remains in a thermal state, the presence of a coherence flow described by $\mathcal{J}_{c}(t)$ seems a puzzling issue at first glance. However, unlike the classical irreversible processes, quantum coherence can flow in a many-body system without being localized in space. The single collision of the qubit S with a projectile pair opens two such delocalized channels through which a micro-current of shared coherence circulates the global three-qubit system. The collision model framework allows us to make a microscopical surgery of $\mathcal{J}_{c}(t)$ focusing on this micro-current. As shown in Fig.~\ref{Fig_Jc}, the initial coherence shared between projectile qubits $B_1$ and $B_2$ drops in $S$-$B_1$ and $S$-$B_2$ joint systems without a localization in any single-qubit subsystem and then returns back to $B_1$-$B_2$ joint system. Although this micro-current disappears when we reset the projectile pair at the end of a single collision, multiple collisions create an average delocalized coherence that flows being shared between the qubit and projectile bath.

A natural question to ask at this point is that why the coherence of the entangled
projectiles does not affect the heat flow by inducing a similar coherence flow?

Although the non-zero values of $\lambda$ and $\mu$ seem to generate quantum coherences in projectile states $\rho_D$ and $\rho_E$ which differ only in the amount in Fig.~\ref{Fig_QBath}~(a)~and~(b), these coherences were found to pertain to different physical processes in our previous studies~\cite{2016_Entropy_Ozgur, 2019_PRE_Ozgur_Multiatom} of qubit baths. The coherence of discordant projectiles was classified as heat-exchange coherence (HEC), whereas the coherence of entangled projectiles was categorized as squeezing coherence. HECs are also known as ``internal''~\cite{AHF_2019_PhysRevResearch_Petruccione} or ``horizontal''~\cite{AutoThermalMach_2019, 2020_PRA_HorizantolCohAndPops, 2020_arXiv_2006_01166} coherences in the literature. Please see Ref.~\cite{2020_AsliReview} for the details.

On the other hand, two entangled projectile pairs share a HEC of $2 \mu^2$ in the state $\rho_E \otimes \rho_E$. As a matter of fact, when we extend the collision block to include two projectile pairs, target qubit's temperature acquires a $\mu^2$ dependence after collective collisions with entangled projectiles (Fig.~\ref{Fig_4QE}~(a)). Also, as shown in Fig.~\ref{Fig_4QE}~(b), such energy-preserving collisions can generate an anomalous heat current which depends on $\mu^2$ in the case of $\Delta \beta = 0$.

Based on these observations, we surmise that the mutual induction of heat and coherence currents originates from energy-preserving collisions with locally thermal projectiles as long as their quantum correlations are associated with HECs.

\subsection{Quantum Thermocoherent Onsager Relations} \label{Sec_Onsager}

$- \Delta C/T$ appears in Eqs.~(\ref{Eq_Js2})~and~(\ref{Eq_Jc}) as if quantum coherence were a thermodynamic potential, like the chemical potential $\mu_e$ in the thermoelectric effect~\cite{2015_ThElec},
\begin{eqnarray}
\mathcal{J}_Q &=& L_{11} \nabla (1/T) + L_{12} \nabla (-\mu_e/T) , \label{Eq_TE1}\\
\mathcal{J}_N &=& L_{21} \nabla (1/T) + L_{22} \nabla (-\mu_e/T) , \label{Eq_TE2}
\end{eqnarray}
where $\mathcal{J}_Q$ and $\mathcal{J}_N$ are, respectively, heat and particle currents. The entropy production rate in thermoelectric phenomena can be written in terms of these currents and the corresponding thermodynamic forces,
\begin{equation}
\Pi^{T\!E} = \mathcal{J}_Q \cdot \nabla (1/T) + \nabla (-\mu_e/T) \cdot \mathcal{J}_N ,
\end{equation}
and the Onsager relation~\cite{1931_Onsager} requires that $L_{12} = L_{21}$.

However, our single-bath model is limited to the transient regime, and it is necessary for us to explore the case of multiple baths for a more practical and direct analogy to steady-state thermoelectricity. For that aim, we include a second bath in the master equation~(\ref{Eq_meq2}). Assume that it consists of projectiles in state~(\ref{Eq_rho_2QD}), with $\beta_{B^\prime}$ and $\lambda^\prime$. When there is no difference either in the strength or in the duration of collisions, the steady-state heat flow from the first bath to the qubit can be written in the high-temperature limit as
\begin{equation}\label{Eq_J1SS}
\mathcal{J}_{h} \approx L_{hh} \Delta \beta - L_{hc} \, \beta_{S|BB^\prime}^\infty \, \Delta C ,
\end{equation}
where $\Delta \beta = (\beta_{B^\prime} - \beta_B)$, $\Delta C = 2 \, \alpha(\lambda^\prime - \lambda)$, $\beta_{S|BB^\prime}^\infty=(\beta_B + \beta_{B^\prime})/(2\,(1 + \alpha \lambda + \alpha \lambda^\prime))$ is the steady-state temperature, and $L_{hh} = L_{hc} = \tilde{L}_{hc}/2$. The first law of thermodynamics is satisfied in the form
$\mathcal{J}_{h} + \mathcal{J}_{h^\prime} = 0$.

The second law in Eq.~(\ref{Eq_EntropyPro}), for the case of double baths~\cite{2019_PRL_Esposito_TDofInfoFlow, 2019_PRL_Esposito_PiWithBathCoh}, reads
\begin{eqnarray}\begin{aligned}\label{Eq_EntropyProSS}
\Pi(\infty) &= \dot{S}_{\mathrm{vN}}[\rho_{S|BB^\prime}^\infty] + \mathrm{tr}[\mathcal{D}_h(\rho_{S|BB^\prime}^\infty) \ln(\rho_{S|B}^\infty)] \\
&\qquad + \mathrm{tr}[\mathcal{D}_{h^\prime}(\rho_{S|BB^\prime}^\infty) \ln(\rho_{S|B^\prime}^\infty)] \\
&= - \mathcal{J}_{h} \beta_{S|B}^\infty - \mathcal{J}_{h^\prime} \beta_{S|B^\prime}^\infty ,
\end{aligned}\end{eqnarray}
where the effective temperature $\beta_{S|B}^\infty$ ($\beta_{S|B^\prime}^\infty$) is the thermalization temperature attained only in the presence of the first (second) bath. This equation says that entropy is generated only in the baths. When we write it as the bilinear product of generalized forces and generalized fluxes in the high-temperature limit, we get
\begin{equation}\label{Eq_EntropyProSS1}
\Pi(\infty) \approx \Delta \beta \, \mathcal{J}_{h}  - \big(\beta_{S|B}^\infty \, C^\prime - \beta_{S|B^\prime}^\infty \, C\big) \, \mathcal{J}_c ,
\end{equation}
where the coherence current $\mathcal{J}_{c}$ is identified as
\begin{equation}\label{Eq_JcSS}
-\mathcal{J}_{c} \equiv L_{ch} \Delta \beta - L_{cc} \, \beta_{S|BB^\prime}^\infty \, \Delta C ,
\end{equation}
with the coefficients $L_{cc} = L_{ch} = L_{hc}$. The minus sign preceding $\mathcal{J}_c$ means that the heat and coherence flow in opposite directions.

Equations~(\ref{Eq_J1SS})~and~(\ref{Eq_JcSS}) bear a striking similarity to the thermoelectric effect described in Eqs.~(\ref{Eq_TE1})~and~(\ref{Eq_TE2}). This leads us to introduce the term ``\textit{thermocoherent effect}'' to dub the present phenomena. But there is only a formal analogy between the particle current $\mathcal{J}_N$ in Eq.~(\ref{Eq_TE2}) and the coherent current $\mathcal{J}_{c}$ which has no fundamental analog in classical physics.

The explicit system under consideration is not artificial, but is a faithful representation of an actual natural thermalization scheme. Also, the mutual induction of heat and coherence currents revealed here is not restricted to the high-temperature regime. Please note that the time-independent amplitude $\mathcal{J}_0$ that is given for more general temperatures in (\ref{Eq_Js0}) is split into two terms that depend on $\Delta \beta$ and $\Delta C$. The same is true for steady-state heat and coherence flows from the first bath to the qubit at any temperature. However, the linear dependence of the flows on the potentials $\Delta \beta$ and $-\beta\Delta C$ only appears in the high-temperature limit. Therefore, the quantum thermocoherent Onsager relations (\ref{Eq_J1SS})~and~(\ref{Eq_JcSS}) are only applicable in this regime.

\section{APPLICATIONS AND POTENTIAL EXPERIMENTS}\label{Sec_Apps}

Discordant states that we found to yield thermocoherent effects can be produced directly by thermal resources~\cite{2017_PRA_Ozgur}. It is experimentally produced in nuclear magnetic resonance (NMR) experiments with CHCl$_3$ molecules~\cite{AHF_2019_NatCommun_Lutz} and $^{187}$Yb$^+$ ion trap simulations~\cite{AHF_2020_arXiv_IonTraps} to demonstrate heat flow reversals. Thermocoherent Onsager relations provide a natural explanation for such heat flow reversals, e.g., it is straightforward to find some $\tilde{L}_{jk}$ and $\Delta C$ values for which $T_S(\infty) < T_S(0) < T_B$ or $T_B < T_S < T_S(\infty)$. Also, a more comprehensive set of thermocoherent effects that we explored systematically here could be examined in similar setups.

The thermocoherent Peltier effect can be exploited to inject quantum coherence into a region to make compact nonthermal cooling or heating environments, where injecting heat would be inefficient. Heat transfer in a quantum or nanoscale can be controlled and induced using well-focused coherent lasers. Such applications may pave the way for a temperature manipulation over different regions within a single molecule. A promising example can be polymerase chain reaction (PCR) technology, where Peltier devices are commonly used to control such temperature-sensitive biochemical reactions.

The thermocoherent Seebeck effect may have implications for coherent current generation by a thermal gradient. According to Eqs.~(\ref{Eq_Jc})~and~(\ref{Eq_JcSS}), the temperature difference can produce a flow of coherence even when $\Delta C = 0$. Such a feat allows the thermal production of quantum coherence and correlations either inside a single bath~\cite{2017_PRA_Ozgur} or between different baths~\cite{2019_PRL_Esposito_PiWithBathCoh}, which has wide-ranging ramifications not only for quantum computers and heat engines, but also for phononic devices~\cite{2012_Phononics}. Thermal diodes and transistors that can rectify heat flows on the micro- and nanoscales~\cite{2012_Phononics} may be particularly used in the thermocoherent preparation of qubits for information processing and thermodynamic tasks. Hence, our findings may
motivate further studies to design quantum technologies based on phononics rather than electronics.

Optimization of thermocoherent coefficients can be used to design ``dicoherent materials" to preserve coherence in a reservoir. Here we showed simple optimization possibilities in terms of the collision time and collectivity. Collective collisions in the short time limit result in thermocoherent coefficients proportional to $8 \, J^2 \tau^2$, whereas $L_{ch}$ and $L_{hc}$ are found four times less for sequential collisions. Hence, the thermocoherent resistivity of the target atom increases with a decrease in collision collectivity, making sequential collisions a better strategy to protect the quantumness of projectiles. On the other hand, experimental setups that allow simultaneous collisions are more suitable to utilize the interconversion between heat and quantum coherence for specific purposes.

Figure~\ref{Fig_Jc} shows that the optimization of thermocoherent coefficients to design dicoherent materials is neither sensitive to the temperature of the projectiles nor to the temperature of the qubit. In all four panels, the magnitude of the coherence and heat currents become zero at the value of $\tau$ that we pinpoint by the second vertical gray line. This corresponds to the perfect preservation of the initial coherence and temperature of the projectiles. The magnitude of the currents reaches its maximum when we halve this value of $\tau$. This maximum value, and therefore the thermocoherent effect, can be increased or decreased by adjusting the temperatures.

The thermocoherent phenomena promise improved control over chemical reactions and nanofabrication techniques. Onsager, in his pioneering paper~\cite{1931_Onsager}, investigated a chemical monomolecular triangle reaction $A \rightleftarrows B \rightleftarrows C \rightleftarrows A$ to exemplify his reciprocity relations. Although this investigation was restricted to the currents for population fractions, a molecule can exist in the quantum coherent superposition of two configurations, e.g., configurations $A$ and $B$ can share quantum coherence and correlations like the qubit pairs forming our projectiles. According to our results, such coherences generate new flows from or to configuration $C$, which can coherently change the chemical equilibrium. Searching for such mechanisms in biological systems or utilizing them for nanofabrication applications can be envisioned.

\section{CONCLUSIONS} \label{Sec_Conclusions}

This paper follows Lord Rayleigh and Onsager's footsteps to systematically establish quantum Onsager relations and associated thermocoherent effects in a simple system where a qubit is bombarded from a side by two-qubit projectiles. We allow for arbitrary collision times and consider a set of quantum states of the projectiles with classical and quantum correlations. In addition, sequential and collective collision scenarios are examined. We find that entangled and discordant projectiles contribute to heat exchange provided that the quantum correlations shared between them are associated with so-called heat exchange coherences. Proper identification of the discord contribution to energy exchange as heat is given by deriving the heat conduction equation, developed by the Fokker-Planck equation out of the master equation. By calculating the heat current analytically and deducing the coherence current from the entropy production rate, we showed that quantum Onsager relations can be derived both in transient and in steady-state regimes; coherent Peltier and coherent Seebeck effects can also be predicted. Thermocoherent effects can be further optimized by utilizing collision times and collectivity.

We discussed a set of applications of these effects, such as coherence-based cooling or heating, coherent heat management at the quantum scale, coherence protection by dicoherent materials, and thermocoherent control of molecular processes. Proof of principle experiments can be proposed in existing nuclear magnetic resonance systems, trapped ions, and circuit or cavity QED setups.

\textbf{Note added.} Recently, we learned that similar reciprocal relations were also developed between heat and squeezing fluxes in Ref.~\cite{2020_arXiv_2011_04560}.

\begin{acknowledgments}

O.P. acknowledges support by the Scientific and Technological Research Council of Turkey (T\"{U}B\.{I}TAK), Grant No. (120F089).

\end{acknowledgments}

\appendix

\section{THE MASTER EQUATION} \label{Appendix::MEq}

The typical micromaser or collision model master equations that rely on short interaction time approximation~\cite{Maser1986, ScullyZubairy1997, MeystreSargent2007} are one of the key elements of the theory of quantum thermalization~\cite{2016_Entropy_Ozgur, 2019_PRE_Ozgur_Multiatom, QThermalization2006, QThermalization2010, QThermalization2014}. The derivation of the master equation here generalizes them to the arbitrary collision times.

\subsection{Derivation}\label{Appendix::MEq1}

After a collision with the qubit, the projectile is discarded for a faithful representation of the Rayleigh problem where another collision with the same pair is a rare event (cf.~Fig.~\ref{Fig_Rayleigh}). Each collision is therefore with a fresh projectile so that we can write a factorized joint qubit-projectile state $\rho(t_j)=\rho_S(t_j) \otimes \rho_B$ at the arrival time $t_j$, where $\rho_B$ is the initial state of the projectile, described in Sec.~\ref{Sec_ProjectileStates}.

Assuming random arrival times, typical for gas of particles, and introducing a probability of $p \, \delta t$ for a collision to occur in a time interval $(t,t+\delta t)$, where $p$ is the rate of the Poisson point process, the evolution of the joint state $\tilde{\rho}$ in the interaction picture can be written as
\begin{equation}\label{Eq_rho_Mix}
\tilde{\rho}(t + \delta t) = p \, \delta t \, U(\tau) \tilde{\rho}(t) U^\dagger(\tau) + (1 - p \, \delta t) \tilde{\rho}(t).
\end{equation}

Taking the limit $\delta t \rightarrow 0$ and tracing out the projectile degrees of freedom, we end up with
\begin{eqnarray}\begin{aligned}\label{Eq_meq}
\frac{\mathrm{d}}{\mathrm{d}t}\tilde{\rho}_S(t) = \mathrm{tr}_B[p \, U(\tau) \tilde{\rho}(t) U^\dagger(\tau) - p \, \tilde{\rho}(t) ].
\end{aligned}\end{eqnarray}
Here, $\tilde{\rho}(t) = \tilde{\rho}_S(t) \otimes \tilde{\rho}_B$, where $\tilde{\rho}_S(t)$ and $\tilde{\rho}_B$ are the interaction picture states of the qubit and the projectile, respectively.

\subsection{Lindblad Form}\label{Appendix::MEq2}

Now, we decompose the master equation above into the two dissipative terms for the discordant projectiles,
\begin{eqnarray}\label{Eq_meq2}
\frac{\mathrm{d}}{\mathrm{d}t}\rho_S(t) = - \frac{\text{i}}{\hbar} [H_S,\rho_S] + \mathcal{D}_h(\rho_S)  + \mathcal{D}_d(\rho_S),
\end{eqnarray}
where the dissipators can be written in the following Lindblad forms~\cite{BreuerAndPetruccione-2002}:
\begin{subequations}\label{Eq_dissips}
\begin{align}
\mathcal{D}_h(\rho_S) &= \sum_{j=1}^2 \kappa_{j} \big(A_j \rho_S A^\dagger_j -\frac{1}{2}\{A^\dagger_j A_j, \rho_S\}\big) , \label{Eq_dissipHEC} \\
\mathcal{D}_d(\rho_S) &= \sum_{j,k=1}^2 \bar{\kappa}_{j,k} \big(\bar{A}_j \rho_S \bar{A}^\dagger_k -\frac{1}{2}\{\bar{A}^\dagger_k \bar{A}_j, \rho_S\}\big) , \label{Eq_dissipDecoh}
\end{align}
\end{subequations}
where $A_1 = \sigma^-$, $A_2 = \sigma^+$, $\bar{A}_1 = \sigma^-\sigma^+$, and $\bar{A}_2 = \sigma^+\sigma^-$.

$\mathcal{D}_d(\rho_S)$ can be cast into the simple form
\begin{equation}
\mathcal{D}_d(\tilde{\rho}_S) = \begin{pmatrix} 0 & - c  {\tilde{\rho}_S}_{1,2} \\
- c  {\tilde{\rho}_S}_{2,1} & 0 \end{pmatrix} ,
\end{equation}
where $c := \bar{\kappa}_{1,2}-(\bar{\kappa}_{1,1}+\bar{\kappa}_{2,2})/2$. We find $c=\sin^4(J \tau)/2\geq 0$ and $c=2 (p_g^2 + p_e^2) \sin^4(\sqrt{2}J \tau)\geq 0$ for sequential and collective collisions, respectively. Moreover, $\mathrm{tr}[H_S\,\mathcal{D}_d(\rho_S)] = 0$. Accordingly, $\mathcal{D}_d(\rho_S)$
describes energy preserving dephasing collisions of projectiles with the qubit, which does not contribute to the heat flow.

$\mathcal{D}_h(\rho_S)$ describes two competing processes associated with favorable and unfavorable collisions~\cite{rayleigh1891}, following the terminology of Lord Rayleigh. In contrast to
dephasing collisions, these processes result in energy exchange between the projectiles and the qubit in the form of heat, as we will further justify below by calculating a heat conduction equation for the system. We find the dissipation rates $\kappa_{1}=\kappa_{2}=(p_{g/e} + \alpha \lambda)\,\eta/\alpha$. Favorable (unfavorable) collisions result in a transition from $|e\rangle \rightarrow |g\rangle$ ($|g\rangle \rightarrow |e\rangle$) at a rate proportional to $p_g$ ($p_e$).
Under the detailed balance of favorable and unfavorable collisions, the competition ends with
an equilibrium state that can be described by an empirical temperature $T_{S}(\infty)$ (cf.~Eq.~(\ref{Eq_T_Sinf})). $T_{S}(\infty)$ can be different than the local temperature of projectile qubits due to the contribution of $\alpha \lambda$ terms in $\kappa_{1/2}$.

\section{THE FOKKER-PLANCK EQUATION} \label{Appendix::FPEq}

Remarkably, identification of the energy flow described by Eq.~(\ref{Eq_HeatCurrent}) as heat is not trivial in quantum thermodynamics, especially in the case of nonthermal baths. A variation of the state may not always increase entropy, and worklike contributions in Eq.~(\ref{Eq_HeatCurrent}) need to be removed to properly define the heat flow~\cite{2019_EPJSpecTop_Kurizki}.

To ensure the projectile state's discordant contribution can be identified as a genuine heat flow for proper thermocoherent description, we further examine the generalized Fokker-Planck equation using Haken's representation~\cite{Carmichael-1999}. Assuming the initial thermal condition for the qubit, we can exactly drop $\mathcal{D}_d$ out of the open system dynamics. The phase-space distribution $P(\upsilon, \upsilon^*, m,t)$ then satisfies
\begin{eqnarray} \label{Eq_FP}
\frac{\partial P}{\partial t} = L\Big(\upsilon, \upsilon^*, m, \frac{\partial}{\partial \upsilon}, \frac{\partial}{\partial \upsilon^*}, \frac{\partial}{\partial m}, \Big) P ,
\end{eqnarray}
which yields
\begin{eqnarray} \begin{aligned}
P &= \Gamma_2 \, \delta(m-1)\, \delta^{(2)}(v) + \Gamma_1 \, \delta(m+1) \, \delta^{(2)}(v) \\
&\qquad + \Gamma_2 \, \delta(m+1) \frac{\partial^2}{\partial \upsilon \partial \upsilon^*} \delta^{(2)}(v) .
\end{aligned}\end{eqnarray}
Here $\Gamma_{1/2} = \langle g/e | \rho_S(t)| g/e \rangle$, $m$ and $\upsilon$ are, respectively, the inversion and polarization variables, and the differential operator $L$ is equal to
\begin{eqnarray}\begin{aligned}
L &= \mathrm{i} \, (E_e - E_g) \Big(\frac{\partial}{\partial \upsilon} \upsilon - \frac{\partial}{\partial \upsilon^*} \upsilon^*\Big)\\
&\;+(p_g + \alpha \lambda)\frac{\eta}{2 \alpha}\Big[(e^{2\frac{\partial}{\partial m}}-1)(1+m)+\frac{\partial}{\partial \upsilon} \upsilon+\frac{\partial}{\partial \upsilon^*} \upsilon^*\Big]\\
&\;+(p_e + \alpha \lambda)\frac{\eta}{2 \alpha}\Big[(e^{-2\frac{\partial}{\partial m}}-1)(1-m)+2\frac{\partial^2}{\partial \upsilon \, \partial \upsilon^{*}}\\
&\;+2\Big(e^{-2\frac{\partial}{\partial m}}+\frac{\partial^2}{\partial \upsilon \, \partial \upsilon^{*}}-\frac{1}{2}\Big)\Big(\frac{\partial}{\partial \upsilon} \upsilon + \frac{\partial}{\partial \upsilon^*} \upsilon^*\Big) \\
&\;+\frac{\partial^4}{\partial \upsilon^2 \partial \upsilon^{*2}}e^{2\frac{\partial}{\partial m}}(1+m)\Big] .
\end{aligned}\end{eqnarray}

We can identify the heat conduction terms in Eq.~(\ref{Eq_FP}) by neglecting the drift terms in $L$, as well as the derivatives beyond the second order. The generalized Fokker-Planck equation then turns out to be the heat equation
\begin{eqnarray} \label{Eq_FP2}
\frac{\partial P}{\partial t} &= &\frac{\eta}{\alpha}\Big[(p_{e} + \alpha \lambda) \Big(\frac{\partial^2}{\partial \upsilon \, \partial \upsilon^{*}}- 2 \frac{\partial^2}{\partial m \, \partial \upsilon} \upsilon - 2 \frac{\partial^2}{\partial m \, \partial \upsilon^*} \upsilon^*\Big) \nonumber \\&&\qquad+ \big(1 + m (p_g - p_e) + 2\, \alpha \lambda \big)\frac{\partial^2}{\partial m^2} \Big] P ,
\end{eqnarray}
where each thermal conductivity coefficient can be explicitly split into a purely thermal population term and a quantum coherent $\alpha \lambda$ term. This quantum heat conduction equation justifies that initial quantum coherence of projectile qubits yielding discordant correlations contributes to the genuine heat flow.

\clearpage


\begin{thebibliography}{56}%
\makeatletter
\providecommand \@ifxundefined [1]{%
 \@ifx{#1\undefined}
}%
\providecommand \@ifnum [1]{%
 \ifnum #1\expandafter \@firstoftwo
 \else \expandafter \@secondoftwo
 \fi
}%
\providecommand \@ifx [1]{%
 \ifx #1\expandafter \@firstoftwo
 \else \expandafter \@secondoftwo
 \fi
}%
\providecommand \natexlab [1]{#1}%
\providecommand \enquote  [1]{``#1''}%
\providecommand \bibnamefont  [1]{#1}%
\providecommand \bibfnamefont [1]{#1}%
\providecommand \citenamefont [1]{#1}%
\providecommand \href@noop [0]{\@secondoftwo}%
\providecommand \href [0]{\begingroup \@sanitize@url \@href}%
\providecommand \@href[1]{\@@startlink{#1}\@@href}%
\providecommand \@@href[1]{\endgroup#1\@@endlink}%
\providecommand \@sanitize@url [0]{\catcode `\\12\catcode `\$12\catcode
  `\&12\catcode `\#12\catcode `\^12\catcode `\_12\catcode `\%12\relax}%
\providecommand \@@startlink[1]{}%
\providecommand \@@endlink[0]{}%
\providecommand \url  [0]{\begingroup\@sanitize@url \@url }%
\providecommand \@url [1]{\endgroup\@href {#1}{\urlprefix }}%
\providecommand \urlprefix  [0]{URL }%
\providecommand \Eprint [0]{\href }%
\providecommand \doibase [0]{https://doi.org/}%
\providecommand \selectlanguage [0]{\@gobble}%
\providecommand \bibinfo  [0]{\@secondoftwo}%
\providecommand \bibfield  [0]{\@secondoftwo}%
\providecommand \translation [1]{[#1]}%
\providecommand \BibitemOpen [0]{}%
\providecommand \bibitemStop [0]{}%
\providecommand \bibitemNoStop [0]{.\EOS\space}%
\providecommand \EOS [0]{\spacefactor3000\relax}%
\providecommand \BibitemShut  [1]{\csname bibitem#1\endcsname}%
\let\auto@bib@innerbib\@empty
\bibitem [{\citenamefont {Strutt~(3rd Baron~Rayleigh)}(1891)}]{rayleigh1891}%
  \BibitemOpen
  \bibfield  {author} {\bibinfo {author} {\bibfnamefont {J.~W.}\ \bibnamefont
  {Strutt~(3rd Baron~Rayleigh)}},\ }\href {\doibase 10.1080/14786449108620207}
  {\bibfield  {journal} {\bibinfo  {journal} {Lond. Edinb. Dubl. Philos. Mag.}\ }\textbf {\bibinfo {volume}
  {32}},\ \bibinfo {pages} {424} (\bibinfo {year} {1891})}\BibitemShut
  {NoStop}%
\bibitem [{\citenamefont {Onsager}(1931)}]{1931_Onsager}%
  \BibitemOpen
  \bibfield  {author} {\bibinfo {author} {\bibfnamefont {L.}~\bibnamefont
  {Onsager}},\ }\bibfield  {title} {\bibinfo {title} {Reciprocal relations in
  irreversible processes. I.},\ }\href {https://doi.org/10.1103/PhysRev.37.405}
  {\bibfield  {journal} {\bibinfo  {journal} {Phys. Rev.}\ }\textbf {\bibinfo
  {volume} {37}},\ \bibinfo {pages} {405} (\bibinfo {year} {1931})}\BibitemShut
  {NoStop}%
\bibitem [{\citenamefont {Behnia}(2015)}]{2015_ThElec}%
  \BibitemOpen
  \bibfield  {author} {\bibinfo {author} {\bibfnamefont {K.}~\bibnamefont
  {Behnia}},\ }\href
  {https://doi.org/10.1093/acprof:oso/9780199697663.001.0001} {\emph {\bibinfo
  {title} {Fundamentals of Thermoelectricity}}},\ \bibinfo {edition} {1st}\
  ed.\ (\bibinfo  {publisher} {Oxford University Press},\ \bibinfo {year}
  {2015})\ Chap.~\bibinfo {chapter} {1}, pp.\ \bibinfo {pages}
  {1--17}\BibitemShut {NoStop}%
\bibitem [{\citenamefont {Scarani}\ \emph {et~al.}(2002)\citenamefont
  {Scarani}, \citenamefont {Ziman}, \citenamefont {\ifmmode \check{S}\else
  \v{S}\fi{}telmachovi\ifmmode~\check{c}\else \v{c}\fi{}}, \citenamefont
  {Gisin},\ and\ \citenamefont {Bu\ifmmode~\check{z}\else
  \v{z}\fi{}ek}}]{2002_PRL_88_097905}%
  \BibitemOpen
  \bibfield  {author} {\bibinfo {author} {\bibfnamefont {V.}~\bibnamefont
  {Scarani}}, \bibinfo {author} {\bibfnamefont {M.}~\bibnamefont {Ziman}},
  \bibinfo {author} {\bibfnamefont {P.}~\bibnamefont {\ifmmode \check{S}\else
  \v{S}\fi{}telmachovi\ifmmode~\check{c}\else \v{c}\fi{}}}, \bibinfo {author}
  {\bibfnamefont {N.}~\bibnamefont {Gisin}},\ and\ \bibinfo {author}
  {\bibfnamefont {V.}~\bibnamefont {Bu\ifmmode~\check{z}\else \v{z}\fi{}ek}},\
  }\bibfield  {title} {\bibinfo {title} {Thermalizing quantum machines:
  Dissipation and entanglement},\ }\href
  {https://doi.org/10.1103/PhysRevLett.88.097905} {\bibfield  {journal}
  {\bibinfo  {journal} {Phys. Rev. Lett.}\ }\textbf {\bibinfo {volume} {88}},\
  \bibinfo {pages} {097905} (\bibinfo {year} {2002})}\BibitemShut {NoStop}%
\bibitem [{\citenamefont {Bernardes}\ \emph {et~al.}(2014)\citenamefont
  {Bernardes}, \citenamefont {Carvalho}, \citenamefont {Monken},\ and\
  \citenamefont {Santos}}]{2014_PRA_90_032111_CB}%
  \BibitemOpen
  \bibfield  {author} {\bibinfo {author} {\bibfnamefont {N.~K.}\ \bibnamefont
  {Bernardes}}, \bibinfo {author} {\bibfnamefont {A.~R.~R.}\ \bibnamefont
  {Carvalho}}, \bibinfo {author} {\bibfnamefont {C.~H.}\ \bibnamefont
  {Monken}},\ and\ \bibinfo {author} {\bibfnamefont {M.~F.}\ \bibnamefont
  {Santos}},\ }\bibfield  {title} {\bibinfo {title} {Environmental correlations
  and Markovian to non-Markovian transitions in collisional models},\ }\href
  {https://doi.org/10.1103/PhysRevA.90.032111} {\bibfield  {journal} {\bibinfo
  {journal} {Phys. Rev. A}\ }\textbf {\bibinfo {volume} {90}},\ \bibinfo
  {pages} {032111} (\bibinfo {year} {2014})}\BibitemShut {NoStop}%
\bibitem [{\citenamefont {Lorenzo}\ \emph {et~al.}(2015)\citenamefont
  {Lorenzo}, \citenamefont {Farace}, \citenamefont {Ciccarello}, \citenamefont
  {Palma},\ and\ \citenamefont {Giovannetti}}]{CM_2015_PRA_Ciccarello}%
  \BibitemOpen
  \bibfield  {author} {\bibinfo {author} {\bibfnamefont {S.}~\bibnamefont
  {Lorenzo}}, \bibinfo {author} {\bibfnamefont {A.}~\bibnamefont {Farace}},
  \bibinfo {author} {\bibfnamefont {F.}~\bibnamefont {Ciccarello}}, \bibinfo
  {author} {\bibfnamefont {G.~M.}\ \bibnamefont {Palma}},\ and\ \bibinfo
  {author} {\bibfnamefont {V.}~\bibnamefont {Giovannetti}},\ }\bibfield
  {title} {\bibinfo {title} {Heat flux and quantum correlations in dissipative
  cascaded systems},\ }\href {https://doi.org/10.1103/PhysRevA.91.022121}
  {\bibfield  {journal} {\bibinfo  {journal} {Phys. Rev. A}\ }\textbf {\bibinfo
  {volume} {91}},\ \bibinfo {pages} {022121} (\bibinfo {year}
  {2015})}\BibitemShut {NoStop}%
\bibitem [{\citenamefont
  {Ciccarello}(2017)}]{CM_2017_QMeasQMetrol_Ciccarello_CMsInQOptics}%
  \BibitemOpen
  \bibfield  {author} {\bibinfo {author} {\bibfnamefont {F.}~\bibnamefont
  {Ciccarello}},\ }\bibfield  {title} {\bibinfo {title} {Collision models in
  quantum optics},\ }\href {https://doi.org/10.1515/qmetro-2017-0007}
  {\bibfield  {journal} {\bibinfo  {journal} {Quantum Meas. Quantum Metrol.}\
  }\textbf {\bibinfo {volume} {4}},\ \bibinfo {pages} {53 } (\bibinfo {year}
  {2017})}\BibitemShut {NoStop}%
\bibitem [{\citenamefont {Filippov}\ \emph {et~al.}(2017)\citenamefont
  {Filippov}, \citenamefont {Piilo}, \citenamefont {Maniscalco},\ and\
  \citenamefont {Ziman}}]{2017_PRA_96_032111_CB}%
  \BibitemOpen
  \bibfield  {author} {\bibinfo {author} {\bibfnamefont {S.~N.}\ \bibnamefont
  {Filippov}}, \bibinfo {author} {\bibfnamefont {J.}~\bibnamefont {Piilo}},
  \bibinfo {author} {\bibfnamefont {S.}~\bibnamefont {Maniscalco}},\ and\
  \bibinfo {author} {\bibfnamefont {M.}~\bibnamefont {Ziman}},\ }\bibfield
  {title} {\bibinfo {title} {Divisibility of quantum dynamical maps and
  collision models},\ }\href {https://doi.org/10.1103/PhysRevA.96.032111}
  {\bibfield  {journal} {\bibinfo  {journal} {Phys. Rev. A}\ }\textbf {\bibinfo
  {volume} {96}},\ \bibinfo {pages} {032111} (\bibinfo {year}
  {2017})}\BibitemShut {NoStop}%
\bibitem [{\citenamefont {Daryanoosh}\ \emph {et~al.}(2018)\citenamefont
  {Daryanoosh}, \citenamefont {Baragiola}, \citenamefont {Guff},\ and\
  \citenamefont {Gilchrist}}]{CM_2018_PRA_EntangledBaths}%
  \BibitemOpen
  \bibfield  {author} {\bibinfo {author} {\bibfnamefont {S.}~\bibnamefont
  {Daryanoosh}}, \bibinfo {author} {\bibfnamefont {B.~Q.}\ \bibnamefont
  {Baragiola}}, \bibinfo {author} {\bibfnamefont {T.}~\bibnamefont {Guff}},\
  and\ \bibinfo {author} {\bibfnamefont {A.}~\bibnamefont {Gilchrist}},\
  }\bibfield  {title} {\bibinfo {title} {Quantum master equations for entangled
  qubit environments},\ }\href {https://doi.org/10.1103/PhysRevA.98.062104}
  {\bibfield  {journal} {\bibinfo  {journal} {Phys. Rev. A}\ }\textbf {\bibinfo
  {volume} {98}},\ \bibinfo {pages} {062104} (\bibinfo {year}
  {2018})}\BibitemShut {NoStop}%
\bibitem [{\citenamefont {De~Chiara}\ and\ \citenamefont
  {Antezza}(2020)}]{2020_arXiv_2006_12848}%
  \BibitemOpen
  \bibfield  {author} {\bibinfo {author} {\bibfnamefont {G.}~\bibnamefont
  {De~Chiara}}\ and\ \bibinfo {author} {\bibfnamefont {M.}~\bibnamefont
  {Antezza}},\ }\bibfield  {title} {\bibinfo {title} {Quantum machines powered
  by correlated baths},\ }\href {https://doi.org/10.1103/PhysRevResearch.2.033315} {\bibfield
  {journal} {\bibinfo  {journal} {Phys. Rev. Res.}\ }\textbf {\bibinfo
  {volume} {2}},\ \bibinfo {pages} {033315} (\bibinfo {year}
  {2020})}\BibitemShut {NoStop}%
\bibitem [{\citenamefont {Brask}\ \emph {et~al.}(2015)\citenamefont {Brask},
  \citenamefont {Haack}, \citenamefont {Brunner},\ and\ \citenamefont
  {Huber}}]{2015_NewJPhys_Brunner_T2Ent}%
  \BibitemOpen
  \bibfield  {author} {\bibinfo {author} {\bibfnamefont {J.~B.}\ \bibnamefont
  {Brask}}, \bibinfo {author} {\bibfnamefont {G.}~\bibnamefont {Haack}},
  \bibinfo {author} {\bibfnamefont {N.}~\bibnamefont {Brunner}},\ and\ \bibinfo
  {author} {\bibfnamefont {M.}~\bibnamefont {Huber}},\ }\bibfield  {title}
  {\bibinfo {title} {Autonomous quantum thermal machine for generating
  steady-state entanglement},\ }\href
  {https://doi.org/10.1088/1367-2630/17/11/113029} {\bibfield  {journal}
  {\bibinfo  {journal} {New J. Phys.}\ }\textbf {\bibinfo {volume} {17}},\
  \bibinfo {pages} {113029} (\bibinfo {year} {2015})}\BibitemShut {NoStop}%
\bibitem [{\citenamefont {Huber}\ \emph {et~al.}(2015)\citenamefont {Huber},
  \citenamefont {Perarnau-Llobet}, \citenamefont {Hovhannisyan}, \citenamefont
  {Skrzypczyk}, \citenamefont {Klöckl}, \citenamefont {Brunner},\ and\
  \citenamefont {Ac{\'{\i}}n}}]{2015_NJP_Brunner_TDCostOfCorrelations}%
  \BibitemOpen
  \bibfield  {author} {\bibinfo {author} {\bibfnamefont {M.}~\bibnamefont
  {Huber}}, \bibinfo {author} {\bibfnamefont {M.}~\bibnamefont
  {Perarnau-Llobet}}, \bibinfo {author} {\bibfnamefont {K.~V.}\ \bibnamefont
  {Hovhannisyan}}, \bibinfo {author} {\bibfnamefont {P.}~\bibnamefont
  {Skrzypczyk}}, \bibinfo {author} {\bibfnamefont {C.}~\bibnamefont {Klöckl}},
  \bibinfo {author} {\bibfnamefont {N.}~\bibnamefont {Brunner}},\ and\ \bibinfo
  {author} {\bibfnamefont {A.}~\bibnamefont {Ac{\'{\i}}n}},\ }\bibfield
  {title} {\bibinfo {title} {Thermodynamic cost of creating correlations},\
  }\href {https://doi.org/10.1088/1367-2630/17/6/065008} {\bibfield  {journal}
  {\bibinfo  {journal} {New J. Phys.}\ }\textbf {\bibinfo {volume} {17}},\
  \bibinfo {pages} {065008} (\bibinfo {year} {2015})}\BibitemShut {NoStop}%
\bibitem [{\citenamefont {\ifmmode~\mbox{\c{C}}\else \c{C}\fi{}akmak}\ \emph
  {et~al.}(2017)\citenamefont {\ifmmode~\mbox{\c{C}}\else \c{C}\fi{}akmak},
  \citenamefont {Manatuly},\ and\ \citenamefont {M\"ustecapl\ifmmode \imath
  \else \i \fi{}o\ifmmode~\breve{g}\else \u{g}\fi{}lu}}]{2017_PRA_Ozgur}%
  \BibitemOpen
  \bibfield  {author} {\bibinfo {author} {\bibfnamefont {B.}~\bibnamefont
  {\ifmmode~\mbox{\c{C}}\else \c{C}\fi{}akmak}}, \bibinfo {author}
  {\bibfnamefont {A.}~\bibnamefont {Manatuly}},\ and\ \bibinfo {author}
  {\bibfnamefont {O.~E.}\ \bibnamefont {M\"ustecapl\ifmmode \imath \else \i
  \fi{}o\ifmmode~\breve{g}\else \u{g}\fi{}lu}},\ }\bibfield  {title} {\bibinfo
  {title} {Thermal production, protection, and heat exchange of quantum
  coherences},\ }\href {https://doi.org/10.1103/PhysRevA.96.032117} {\bibfield
  {journal} {\bibinfo  {journal} {Phys. Rev. A}\ }\textbf {\bibinfo {volume}
  {96}},\ \bibinfo {pages} {032117} (\bibinfo {year} {2017})}\BibitemShut
  {NoStop}%
\bibitem [{\citenamefont {Hu}\ \emph {et~al.}(2018)\citenamefont {Hu},
  \citenamefont {Man},\ and\ \citenamefont
  {Xia}}]{2018_QIP_EntIn2QubitsWith2CommonBaths}%
  \BibitemOpen
  \bibfield  {author} {\bibinfo {author} {\bibfnamefont {L.-Z.}\ \bibnamefont
  {Hu}}, \bibinfo {author} {\bibfnamefont {Z.-X.}\ \bibnamefont {Man}},\ and\
  \bibinfo {author} {\bibfnamefont {Y.-J.}\ \bibnamefont {Xia}},\ }\bibfield
  {title} {\bibinfo {title} {Steady-state entanglement and thermalization of
  coupled qubits in two common heat baths},\ }\href
  {https://doi.org/10.1007/s11128-018-1825-x} {\bibfield  {journal} {\bibinfo
  {journal} {Quantum Inf. Process.}\ }\textbf {\bibinfo {volume} {17}},\
  \bibinfo {pages} {45} (\bibinfo {year} {2018})}\BibitemShut {NoStop}%
\bibitem [{\citenamefont {Tavakoli}\ \emph {et~al.}(2018)\citenamefont
  {Tavakoli}, \citenamefont {Haack}, \citenamefont {Huber}, \citenamefont
  {Brunner},\ and\ \citenamefont {Brask}}]{2018_Quantum_BrunnerT2Ent}%
  \BibitemOpen
  \bibfield  {author} {\bibinfo {author} {\bibfnamefont {A.}~\bibnamefont
  {Tavakoli}}, \bibinfo {author} {\bibfnamefont {G.}~\bibnamefont {Haack}},
  \bibinfo {author} {\bibfnamefont {M.}~\bibnamefont {Huber}}, \bibinfo
  {author} {\bibfnamefont {N.}~\bibnamefont {Brunner}},\ and\ \bibinfo {author}
  {\bibfnamefont {J.~B.}\ \bibnamefont {Brask}},\ }\bibfield  {title} {\bibinfo
  {title} {Heralded generation of maximal entanglement in any dimension via
  incoherent coupling to thermal baths},\ }\href
  {https://doi.org/10.22331/q-2018-06-13-73} {\bibfield  {journal} {\bibinfo
  {journal} {{Quantum}}\ }\textbf {\bibinfo {volume} {2}},\ \bibinfo {pages}
  {73} (\bibinfo {year} {2018})}\BibitemShut {NoStop}%
\bibitem [{\citenamefont {Manzano}\ \emph {et~al.}(2019)\citenamefont
  {Manzano}, \citenamefont {Silva},\ and\ \citenamefont
  {Parrondo}}]{2019_PRE_T2QCoh}%
  \BibitemOpen
  \bibfield  {author} {\bibinfo {author} {\bibfnamefont {G.}~\bibnamefont
  {Manzano}}, \bibinfo {author} {\bibfnamefont {R.}~\bibnamefont {Silva}},\
  and\ \bibinfo {author} {\bibfnamefont {J.~M.~R.}\ \bibnamefont {Parrondo}},\
  }\bibfield  {title} {\bibinfo {title} {Autonomous thermal machine for
  amplification and control of energetic coherence},\ }\href
  {https://doi.org/10.1103/PhysRevE.99.042135} {\bibfield  {journal} {\bibinfo
  {journal} {Phys. Rev. E}\ }\textbf {\bibinfo {volume} {99}},\ \bibinfo
  {pages} {042135} (\bibinfo {year} {2019})}\BibitemShut {NoStop}%
\bibitem [{\citenamefont {Tavakoli}\ \emph {et~al.}(2020)\citenamefont
  {Tavakoli}, \citenamefont {Haack}, \citenamefont {Brunner},\ and\
  \citenamefont {Brask}}]{2020_PRA_Brunner_AutonomousMultipartiteEntEngines}%
  \BibitemOpen
  \bibfield  {author} {\bibinfo {author} {\bibfnamefont {A.}~\bibnamefont
  {Tavakoli}}, \bibinfo {author} {\bibfnamefont {G.}~\bibnamefont {Haack}},
  \bibinfo {author} {\bibfnamefont {N.}~\bibnamefont {Brunner}},\ and\ \bibinfo
  {author} {\bibfnamefont {J.~B.}\ \bibnamefont {Brask}},\ }\bibfield  {title}
  {\bibinfo {title} {Autonomous multipartite entanglement engines},\ }\href
  {https://doi.org/10.1103/PhysRevA.101.012315} {\bibfield  {journal} {\bibinfo
   {journal} {Phys. Rev. A}\ }\textbf {\bibinfo {volume} {101}},\ \bibinfo
  {pages} {012315} (\bibinfo {year} {2020})}\BibitemShut {NoStop}%
\bibitem [{\citenamefont {Partovi}(2008)}]{AHF_2008_PRE_Partovi}%
  \BibitemOpen
  \bibfield  {author} {\bibinfo {author} {\bibfnamefont {M.~H.}\ \bibnamefont
  {Partovi}},\ }\bibfield  {title} {\bibinfo {title} {Entanglement versus
  stosszahlansatz: Disappearance of the thermodynamic arrow in a
  high-correlation environment},\ }\href
  {https://doi.org/10.1103/PhysRevE.77.021110} {\bibfield  {journal} {\bibinfo
  {journal} {Phys. Rev. E}\ }\textbf {\bibinfo {volume} {77}},\ \bibinfo
  {pages} {021110} (\bibinfo {year} {2008})}\BibitemShut {NoStop}%
\bibitem [{\citenamefont {Dillenschneider}\ and\ \citenamefont
  {Lutz}(2009)}]{Lutz2009}%
  \BibitemOpen
  \bibfield  {author} {\bibinfo {author} {\bibfnamefont {R.}~\bibnamefont
  {Dillenschneider}}\ and\ \bibinfo {author} {\bibfnamefont {E.}~\bibnamefont
  {Lutz}},\ }\bibfield  {title} {\bibinfo {title} {Energetics of quantum
  correlations},\ }\href {https://doi.org/10.1209/0295-5075/88/50003}
  {\bibfield  {journal} {\bibinfo  {journal} {Europhys. Lett.}\ }\textbf {\bibinfo {volume}
  {88}},\ \bibinfo {pages} {50003} (\bibinfo {year} {2009})}\BibitemShut
  {NoStop}%
\bibitem [{\citenamefont {Jennings}\ and\ \citenamefont
  {Rudolph}(2010)}]{AHF_2010_PRE_JenningsAndRudolph}%
  \BibitemOpen
  \bibfield  {author} {\bibinfo {author} {\bibfnamefont {D.}~\bibnamefont
  {Jennings}}\ and\ \bibinfo {author} {\bibfnamefont {T.}~\bibnamefont
  {Rudolph}},\ }\bibfield  {title} {\bibinfo {title} {Entanglement and the
  thermodynamic arrow of time},\ }\href
  {https://doi.org/10.1103/PhysRevE.81.061130} {\bibfield  {journal} {\bibinfo
  {journal} {Phys. Rev. E}\ }\textbf {\bibinfo {volume} {81}},\ \bibinfo
  {pages} {061130} (\bibinfo {year} {2010})}\BibitemShut {NoStop}%
\bibitem [{\citenamefont {Da\u{g}}\ \emph {et~al.}(2016)\citenamefont
  {Da\u{g}}, \citenamefont {Niedenzu}, \citenamefont
  {M\"{u}stecapl{\i}o\u{g}lu},\ and\ \citenamefont
  {Kurizki}}]{2016_Entropy_Ozgur}%
  \BibitemOpen
  \bibfield  {author} {\bibinfo {author} {\bibfnamefont {C.}~\bibnamefont
  {Da\u{g}}}, \bibinfo {author} {\bibfnamefont {W.}~\bibnamefont {Niedenzu}},
  \bibinfo {author} {\bibfnamefont {O.}~\bibnamefont
  {M\"{u}stecapl{\i}o\u{g}lu}},\ and\ \bibinfo {author} {\bibfnamefont
  {G.}~\bibnamefont {Kurizki}},\ }\bibfield  {title} {\bibinfo {title}
  {Multiatom quantum coherences in micromasers as fuel for thermal and
  nonthermal machines},\ }\href {https://doi.org/10.3390/e18070244} {\bibfield
  {journal} {\bibinfo  {journal} {Entropy}\ }\textbf {\bibinfo {volume} {18}},\
  \bibinfo {pages} {244} (\bibinfo {year} {2016})}\BibitemShut {NoStop}%
\bibitem [{\citenamefont {Henao}\ and\ \citenamefont
  {Serra}(2018)}]{2018_PRE_RoleOfQCohInHT}%
  \BibitemOpen
  \bibfield  {author} {\bibinfo {author} {\bibfnamefont {I.}~\bibnamefont
  {Henao}}\ and\ \bibinfo {author} {\bibfnamefont {R.~M.}\ \bibnamefont
  {Serra}},\ }\bibfield  {title} {\bibinfo {title} {Role of quantum coherence
  in the thermodynamics of energy transfer},\ }\href
  {https://doi.org/10.1103/PhysRevE.97.062105} {\bibfield  {journal} {\bibinfo
  {journal} {Phys. Rev. E}\ }\textbf {\bibinfo {volume} {97}},\ \bibinfo
  {pages} {062105} (\bibinfo {year} {2018})}\BibitemShut {NoStop}%
\bibitem [{\citenamefont {Latune}\ \emph
  {et~al.}(2019{\natexlab{a}})\citenamefont {Latune}, \citenamefont
  {Sinayskiy},\ and\ \citenamefont
  {Petruccione}}]{AHF_2019_QuantumSciTechnol_Petruccione_AppTs}%
  \BibitemOpen
  \bibfield  {author} {\bibinfo {author} {\bibfnamefont {C.~L.}\ \bibnamefont
  {Latune}}, \bibinfo {author} {\bibfnamefont {I.}~\bibnamefont {Sinayskiy}},\
  and\ \bibinfo {author} {\bibfnamefont {F.}~\bibnamefont {Petruccione}},\
  }\bibfield  {title} {\bibinfo {title} {Apparent temperature: demystifying the
  relation between quantum coherence, correlations, and heat flows},\ }\href
  {https://doi.org/10.1088/2058-9565/aaf5f7} {\bibfield  {journal} {\bibinfo
  {journal} {Quantum Sci. Technol.}\ }\textbf {\bibinfo {volume} {4}},\
  \bibinfo {pages} {025005} (\bibinfo {year} {2019}{\natexlab{a}})}\BibitemShut
  {NoStop}%
\bibitem [{\citenamefont {Latune}\ \emph
  {et~al.}(2019{\natexlab{b}})\citenamefont {Latune}, \citenamefont
  {Sinayskiy},\ and\ \citenamefont {Petruccione}}]{AutoThermalMach_2019}%
  \BibitemOpen
  \bibfield  {author} {\bibinfo {author} {\bibfnamefont {C.~L.}\ \bibnamefont
  {Latune}}, \bibinfo {author} {\bibfnamefont {I.}~\bibnamefont {Sinayskiy}},\
  and\ \bibinfo {author} {\bibfnamefont {F.}~\bibnamefont {Petruccione}},\
  }\bibfield  {title} {\bibinfo {title} {Quantum coherence, many-body
  correlations, and non-thermal effects for autonomous thermal machines},\
  }\href {https://doi.org/10.1038/s41598-019-39300-4} {\bibfield  {journal}
  {\bibinfo  {journal} {Sci. Rep.}\ }\textbf {\bibinfo {volume} {9}},\ \bibinfo
  {pages} {3191} (\bibinfo {year} {2019}{\natexlab{b}})}\BibitemShut {NoStop}%
\bibitem [{\citenamefont {Santos}\ \emph {et~al.}(2019)\citenamefont {Santos},
  \citenamefont {C\'{e}leri}, \citenamefont {Landi},\ and\ \citenamefont
  {Paternostro}}]{2019_npj}%
  \BibitemOpen
  \bibfield  {author} {\bibinfo {author} {\bibfnamefont {J.}~\bibnamefont
  {Santos}}, \bibinfo {author} {\bibfnamefont {L.}~\bibnamefont {C\'{e}leri}},
  \bibinfo {author} {\bibfnamefont {G.}~\bibnamefont {Landi}},\ and\ \bibinfo
  {author} {\bibfnamefont {M.}~\bibnamefont {Paternostro}},\ }\bibfield
  {title} {\bibinfo {title} {The role of quantum coherence in non-equilibrium
  entropy production},\ }\href {https://doi.org/10.1038/s41534-019-0138-y}
  {\bibfield  {journal} {\bibinfo  {journal} {npj Quantum Inf.}\ }\textbf
  {\bibinfo {volume} {5}},\ \bibinfo {pages} {23} (\bibinfo {year}
  {2019})}\BibitemShut {NoStop}%
\bibitem [{\citenamefont {Manatuly}\ \emph {et~al.}(2019)\citenamefont
  {Manatuly}, \citenamefont {Niedenzu}, \citenamefont {Rom\'an-Ancheyta},
  \citenamefont {\ifmmode~\mbox{\c{C}}\else \c{C}\fi{}akmak}, \citenamefont
  {M\"ustecapl\ifmmode \imath \else \i \fi{}o\ifmmode~\breve{g}\else
  \u{g}\fi{}lu},\ and\ \citenamefont {Kurizki}}]{2019_PRE_Ozgur_Multiatom}%
  \BibitemOpen
  \bibfield  {author} {\bibinfo {author} {\bibfnamefont {A.}~\bibnamefont
  {Manatuly}}, \bibinfo {author} {\bibfnamefont {W.}~\bibnamefont {Niedenzu}},
  \bibinfo {author} {\bibfnamefont {R.}~\bibnamefont {Rom\'an-Ancheyta}},
  \bibinfo {author} {\bibfnamefont {B.}~\bibnamefont
  {\ifmmode~\mbox{\c{C}}\else \c{C}\fi{}akmak}}, \bibinfo {author}
  {\bibfnamefont {O.~E.}\ \bibnamefont {M\"ustecapl\ifmmode \imath \else \i
  \fi{}o\ifmmode~\breve{g}\else \u{g}\fi{}lu}},\ and\ \bibinfo {author}
  {\bibfnamefont {G.}~\bibnamefont {Kurizki}},\ }\bibfield  {title} {\bibinfo
  {title} {Collectively enhanced thermalization via multiqubit collisions},\
  }\href {https://doi.org/10.1103/PhysRevE.99.042145} {\bibfield  {journal}
  {\bibinfo  {journal} {Phys. Rev. E}\ }\textbf {\bibinfo {volume} {99}},\
  \bibinfo {pages} {042145} (\bibinfo {year} {2019})}\BibitemShut {NoStop}%
\bibitem [{\citenamefont {Ptaszy\ifmmode~\acute{n}\else \'{n}\fi{}ski}\ and\
  \citenamefont
  {Esposito}(2019{\natexlab{a}})}]{2019_PRL_Esposito_TDofInfoFlow}%
  \BibitemOpen
  \bibfield  {author} {\bibinfo {author} {\bibfnamefont {K.}~\bibnamefont
  {Ptaszy\ifmmode~\acute{n}\else \'{n}\fi{}ski}}\ and\ \bibinfo {author}
  {\bibfnamefont {M.}~\bibnamefont {Esposito}},\ }\bibfield  {title} {\bibinfo
  {title} {Thermodynamics of quantum information flows},\ }\href
  {https://doi.org/10.1103/PhysRevLett.122.150603} {\bibfield  {journal}
  {\bibinfo  {journal} {Phys. Rev. Lett.}\ }\textbf {\bibinfo {volume} {122}},\
  \bibinfo {pages} {150603} (\bibinfo {year} {2019}{\natexlab{a}})}\BibitemShut
  {NoStop}%
\bibitem [{\citenamefont {Ma}\ \emph {et~al.}(2019)\citenamefont {Ma},
  \citenamefont {Zhao}, \citenamefont {Fei},\ and\ \citenamefont
  {Yung}}]{2019_PRA_CohNeeded4HF}%
  \BibitemOpen
  \bibfield  {author} {\bibinfo {author} {\bibfnamefont {T.}~\bibnamefont
  {Ma}}, \bibinfo {author} {\bibfnamefont {M.-J.}\ \bibnamefont {Zhao}},
  \bibinfo {author} {\bibfnamefont {S.-M.}\ \bibnamefont {Fei}},\ and\ \bibinfo
  {author} {\bibfnamefont {M.-H.}\ \bibnamefont {Yung}},\ }\bibfield  {title}
  {\bibinfo {title} {Necessity for quantum coherence of nondegeneracy in energy
  flow},\ }\href {https://doi.org/10.1103/PhysRevA.99.062303} {\bibfield
  {journal} {\bibinfo  {journal} {Phys. Rev. A}\ }\textbf {\bibinfo {volume}
  {99}},\ \bibinfo {pages} {062303} (\bibinfo {year} {2019})}\BibitemShut
  {NoStop}%
\bibitem [{\citenamefont {Micadei}\ \emph {et~al.}(2019)\citenamefont
  {Micadei}, \citenamefont {Peterson}, \citenamefont {Souza},\ and\
  \citenamefont {et~al.}}]{AHF_2019_NatCommun_Lutz}%
  \BibitemOpen
  \bibfield  {author} {\bibinfo {author} {\bibfnamefont {K.}~\bibnamefont
  {Micadei}}, \bibinfo {author} {\bibfnamefont {J.~P.~S.}\ \bibnamefont
  {Peterson}}, \bibinfo {author} {\bibfnamefont {A.~M.}\ \bibnamefont
  {Souza}},\ and\ \bibinfo {author} {\bibnamefont {et~al.}},\ }\bibfield
  {title} {\bibinfo {title} {Reversing the direction of heat flow using quantum
  correlations},\ }\href {https://doi.org/10.1038/s41467-019-10333-7}
  {\bibfield  {journal} {\bibinfo  {journal} {Nat. Commun.}\ }\textbf {\bibinfo
  {volume} {10}},\ \bibinfo {pages} {2456} (\bibinfo {year}
  {2019})}\BibitemShut {NoStop}%
\bibitem [{\citenamefont {Rodr\'{\i}guez-Rosario}\ \emph
  {et~al.}(2019)\citenamefont {Rodr\'{\i}guez-Rosario}, \citenamefont
  {Frauenheim},\ and\ \citenamefont {Aspuru-Guzik}}]{2019_OSID}%
  \BibitemOpen
  \bibfield  {author} {\bibinfo {author} {\bibfnamefont {C.~A.}\ \bibnamefont
  {Rodr\'{\i}guez-Rosario}}, \bibinfo {author} {\bibfnamefont {T.}~\bibnamefont
  {Frauenheim}},\ and\ \bibinfo {author} {\bibfnamefont {A.}~\bibnamefont
  {Aspuru-Guzik}},\ }\bibfield  {title} {\bibinfo {title} {Quantum coherences
  as a thermodynamic potential},\ }\href
  {https://doi.org/10.1142/S1230161219500227} {\bibfield  {journal} {\bibinfo
  {journal} {Open Syst. Inf. Dyn.}\ }\textbf {\bibinfo {volume} {26}},\
  \bibinfo {pages} {1950022} (\bibinfo {year} {2019})}\BibitemShut {NoStop}%
\bibitem [{\citenamefont {Latune}\ \emph
  {et~al.}(2019{\natexlab{c}})\citenamefont {Latune}, \citenamefont
  {Sinayskiy},\ and\ \citenamefont
  {Petruccione}}]{AHF_2019_PhysRevResearch_Petruccione}%
  \BibitemOpen
  \bibfield  {author} {\bibinfo {author} {\bibfnamefont {C.~L.}\ \bibnamefont
  {Latune}}, \bibinfo {author} {\bibfnamefont {I.}~\bibnamefont {Sinayskiy}},\
  and\ \bibinfo {author} {\bibfnamefont {F.}~\bibnamefont {Petruccione}},\
  }\bibfield  {title} {\bibinfo {title} {Heat flow reversals without reversing
  the arrow of time: The role of internal quantum coherences and
  correlations},\ }\href {https://doi.org/10.1103/PhysRevResearch.1.033097}
  {\bibfield  {journal} {\bibinfo  {journal} {Phys. Rev. Res.}\ }\textbf
  {\bibinfo {volume} {1}},\ \bibinfo {pages} {033097} (\bibinfo {year}
  {2019}{\natexlab{c}})}\BibitemShut {NoStop}%
\bibitem [{\citenamefont {Ptaszy\ifmmode~\acute{n}\else \'{n}\fi{}ski}\ and\
  \citenamefont
  {Esposito}(2019{\natexlab{b}})}]{2019_PRL_Esposito_PiWithBathCoh}%
  \BibitemOpen
  \bibfield  {author} {\bibinfo {author} {\bibfnamefont {K.}~\bibnamefont
  {Ptaszy\ifmmode~\acute{n}\else \'{n}\fi{}ski}}\ and\ \bibinfo {author}
  {\bibfnamefont {M.}~\bibnamefont {Esposito}},\ }\bibfield  {title} {\bibinfo
  {title} {Entropy production in open systems: The predominant role of
  intraenvironment correlations},\ }\href
  {https://doi.org/10.1103/PhysRevLett.123.200603} {\bibfield  {journal}
  {\bibinfo  {journal} {Phys. Rev. Lett.}\ }\textbf {\bibinfo {volume} {123}},\
  \bibinfo {pages} {200603} (\bibinfo {year} {2019}{\natexlab{b}})}\BibitemShut
  {NoStop}%
\bibitem [{\citenamefont {Latune}\ \emph
  {et~al.}(2020{\natexlab{a}})\citenamefont {Latune}, \citenamefont
  {Sinayskiy},\ and\ \citenamefont
  {Petruccione}}]{2020_PRA_HorizantolCohAndPops}%
  \BibitemOpen
  \bibfield  {author} {\bibinfo {author} {\bibfnamefont {C.~L.}\ \bibnamefont
  {Latune}}, \bibinfo {author} {\bibfnamefont {I.}~\bibnamefont {Sinayskiy}},\
  and\ \bibinfo {author} {\bibfnamefont {F.}~\bibnamefont {Petruccione}},\
  }\bibfield  {title} {\bibinfo {title} {Negative contributions to entropy
  production induced by quantum coherences},\ }\href
  {https://doi.org/10.1103/PhysRevA.102.042220} {\bibfield  {journal} {\bibinfo
   {journal} {Phys. Rev. A}\ }\textbf {\bibinfo {volume} {102}},\ \bibinfo
  {pages} {042220} (\bibinfo {year} {2020}{\natexlab{a}})}\BibitemShut
  {NoStop}%
\bibitem [{\citenamefont {Medina~Gonz\'alez}\ \emph {et~al.}(2020)\citenamefont
  {Medina~Gonz\'alez}, \citenamefont {Ramos-Prieto},\ and\ \citenamefont
  {Rodr\'{\i}guez-Lara}}]{AHF_2020_arXiv_IonTraps}%
  \BibitemOpen
  \bibfield  {author} {\bibinfo {author} {\bibfnamefont {P.~U.}\ \bibnamefont
  {Medina~Gonz\'alez}}, \bibinfo {author} {\bibfnamefont {I.}~\bibnamefont
  {Ramos-Prieto}},\ and\ \bibinfo {author} {\bibfnamefont {B.~M.}\ \bibnamefont
  {Rodr\'{\i}guez-Lara}},\ }\bibfield  {title} {\bibinfo {title} {Heat-flow
  reversal in a trapped-ion simulator},\ }\href
  {https://doi.org/10.1103/PhysRevA.101.062108} {\bibfield  {journal} {\bibinfo
   {journal} {Phys. Rev. A}\ }\textbf {\bibinfo {volume} {101}},\ \bibinfo
  {pages} {062108} (\bibinfo {year} {2020})}\BibitemShut {NoStop}%
\bibitem [{\citenamefont {Latune}\ \emph
  {et~al.}(2020{\natexlab{b}})\citenamefont {Latune}, \citenamefont
  {Sinayskiy},\ and\ \citenamefont {Petruccione}}]{2020_arXiv_2006_01166}%
  \BibitemOpen
  \bibfield  {author} {\bibinfo {author} {\bibfnamefont {C.~L.}\ \bibnamefont
  {Latune}}, \bibinfo {author} {\bibfnamefont {I.}~\bibnamefont {Sinayskiy}},\
  and\ \bibinfo {author} {\bibfnamefont {F.}~\bibnamefont {Petruccione}},\
  }\bibfield  {title} {\bibinfo {title} {Roles of quantum coherences in thermal
  machines},\ }\href {https://doi.org/10.1140/epjs/s11734-021-00085-1} {\bibfield  {journal}
  {\bibinfo  {journal} {Euro. Phys. J. Spec. Top.}\ } (\bibinfo {year}
  {2021}{\natexlab{b}})}\BibitemShut {NoStop}%
\bibitem [{\citenamefont {Baumgratz}\ \emph {et~al.}(2014)\citenamefont
  {Baumgratz}, \citenamefont {Cramer},\ and\ \citenamefont
  {Plenio}}]{Plenio-2014}%
  \BibitemOpen
  \bibfield  {author} {\bibinfo {author} {\bibfnamefont {T.}~\bibnamefont
  {Baumgratz}}, \bibinfo {author} {\bibfnamefont {M.}~\bibnamefont {Cramer}},\
  and\ \bibinfo {author} {\bibfnamefont {M.~B.}\ \bibnamefont {Plenio}},\
  }\bibfield  {title} {\bibinfo {title} {Quantifying coherence},\ }\href
  {https://doi.org/10.1103/PhysRevLett.113.140401} {\bibfield  {journal}
  {\bibinfo  {journal} {Phys. Rev. Lett.}\ }\textbf {\bibinfo {volume} {113}},\
  \bibinfo {pages} {140401} (\bibinfo {year} {2014})}\BibitemShut {NoStop}%
\bibitem [{\citenamefont {Henderson}\ and\ \citenamefont
  {Vedral}(2001)}]{Vedral-2001}%
  \BibitemOpen
  \bibfield  {author} {\bibinfo {author} {\bibfnamefont {L.}~\bibnamefont
  {Henderson}}\ and\ \bibinfo {author} {\bibfnamefont {V.}~\bibnamefont
  {Vedral}},\ }\bibfield  {title} {\bibinfo {title} {Classical, quantum and
  total correlations},\ }\href {https://doi.org/10.1088/0305-4470/34/35/315}
  {\bibfield  {journal} {\bibinfo  {journal} {J. Phys. A: Math. Gen.}\ }\textbf
  {\bibinfo {volume} {34}},\ \bibinfo {pages} {6899} (\bibinfo {year}
  {2001})}\BibitemShut {NoStop}%
\bibitem [{\citenamefont {Ollivier}\ and\ \citenamefont
  {Zurek}(2001)}]{Zurek-2002}%
  \BibitemOpen
  \bibfield  {author} {\bibinfo {author} {\bibfnamefont {H.}~\bibnamefont
  {Ollivier}}\ and\ \bibinfo {author} {\bibfnamefont {W.~H.}\ \bibnamefont
  {Zurek}},\ }\bibfield  {title} {\bibinfo {title} {Quantum discord: A measure
  of the quantumness of correlations},\ }\href
  {https://doi.org/10.1103/PhysRevLett.88.017901} {\bibfield  {journal}
  {\bibinfo  {journal} {Phys. Rev. Lett.}\ }\textbf {\bibinfo {volume} {88}},\
  \bibinfo {pages} {017901} (\bibinfo {year} {2001})}\BibitemShut {NoStop}%
\bibitem [{\citenamefont {Bennett}\ \emph {et~al.}(1996)\citenamefont
  {Bennett}, \citenamefont {DiVincenzo}, \citenamefont {Smolin},\ and\
  \citenamefont {Wootters}}]{Wooters-1996}%
  \BibitemOpen
  \bibfield  {author} {\bibinfo {author} {\bibfnamefont {C.~H.}\ \bibnamefont
  {Bennett}}, \bibinfo {author} {\bibfnamefont {D.~P.}\ \bibnamefont
  {DiVincenzo}}, \bibinfo {author} {\bibfnamefont {J.~A.}\ \bibnamefont
  {Smolin}},\ and\ \bibinfo {author} {\bibfnamefont {W.~K.}\ \bibnamefont
  {Wootters}},\ }\bibfield  {title} {\bibinfo {title} {Mixed-state entanglement
  and quantum error correction},\ }\href
  {https://doi.org/10.1103/PhysRevA.54.3824} {\bibfield  {journal} {\bibinfo
  {journal} {Phys. Rev. A}\ }\textbf {\bibinfo {volume} {54}},\ \bibinfo
  {pages} {3824} (\bibinfo {year} {1996})}\BibitemShut {NoStop}%
\bibitem [{\citenamefont {Ma}\ \emph {et~al.}(2016)\citenamefont {Ma},
  \citenamefont {Yadin}, \citenamefont {Girolami}, \citenamefont {Vedral},\
  and\ \citenamefont {Gu}}]{Coh2Discord}%
  \BibitemOpen
  \bibfield  {author} {\bibinfo {author} {\bibfnamefont {J.}~\bibnamefont
  {Ma}}, \bibinfo {author} {\bibfnamefont {B.}~\bibnamefont {Yadin}}, \bibinfo
  {author} {\bibfnamefont {D.}~\bibnamefont {Girolami}}, \bibinfo {author}
  {\bibfnamefont {V.}~\bibnamefont {Vedral}},\ and\ \bibinfo {author}
  {\bibfnamefont {M.}~\bibnamefont {Gu}},\ }\bibfield  {title} {\bibinfo
  {title} {Converting coherence to quantum correlations},\ }\href
  {https://doi.org/10.1103/PhysRevLett.116.160407} {\bibfield  {journal}
  {\bibinfo  {journal} {Phys. Rev. Lett.}\ }\textbf {\bibinfo {volume} {116}},\
  \bibinfo {pages} {160407} (\bibinfo {year} {2016})}\BibitemShut {NoStop}%
\bibitem [{\citenamefont {Chitambar}\ and\ \citenamefont
  {Gour}(2019)}]{QRT_2019}%
  \BibitemOpen
  \bibfield  {author} {\bibinfo {author} {\bibfnamefont {E.}~\bibnamefont
  {Chitambar}}\ and\ \bibinfo {author} {\bibfnamefont {G.}~\bibnamefont
  {Gour}},\ }\bibfield  {title} {\bibinfo {title} {Quantum resource theories},\
  }\href {https://doi.org/10.1103/RevModPhys.91.025001} {\bibfield  {journal}
  {\bibinfo  {journal} {Rev. Mod. Phys.}\ }\textbf {\bibinfo {volume} {91}},\
  \bibinfo {pages} {025001} (\bibinfo {year} {2019})}\BibitemShut {NoStop}%
\bibitem [{\citenamefont {Alicki}\ and\ \citenamefont
  {Kosloff}(2019)}]{HeatCurrent}%
  \BibitemOpen
  \bibfield  {author} {\bibinfo {author} {\bibfnamefont {R.}~\bibnamefont
  {Alicki}}\ and\ \bibinfo {author} {\bibfnamefont {R.}~\bibnamefont
  {Kosloff}},\ }\bibinfo {title} {Iintroduction to quantum thermodynamics:
  History and prospects},\ in\ \href
  {https://doi.org/doi.org/10.1007/978-3-319-99046-0_1} {\emph {\bibinfo
  {booktitle} {Thermodynamics in the Quantum Regime}}},\ \bibinfo {series}
  {Fundamental Theories of Physics}, Vol.\ \bibinfo {volume} {195},\ \bibinfo
  {editor} {edited by\ \bibinfo {editor} {\bibfnamefont {F.}~\bibnamefont
  {Binder}}, \bibinfo {editor} {\bibfnamefont {L.}~\bibnamefont {Correa}},
  \bibinfo {editor} {\bibfnamefont {C.}~\bibnamefont {Gogolin}}, \bibinfo
  {editor} {\bibfnamefont {J.}~\bibnamefont {Anders}},\ and\ \bibinfo {editor}
  {\bibfnamefont {G.}~\bibnamefont {Adesso}}}\ (\bibinfo  {publisher} {Springer
  Nature},\ \bibinfo {address} {Cham},\ \bibinfo {year} {2019})\ Chap.~\bibinfo
  {chapter} {1}, pp.\ \bibinfo {pages} {1--33},\ \bibinfo {edition} {1st}\
  ed.\BibitemShut {Stop}%
\bibitem [{\citenamefont {Landi}\ and\ \citenamefont
  {Paternostro}(2020)}]{2020_arXiv_2009_07668}%
  \BibitemOpen
  \bibfield  {author} {\bibinfo {author} {\bibfnamefont {G.}~\bibnamefont
  {Landi}}\ and\ \bibinfo {author} {\bibfnamefont {M.}~\bibnamefont
  {Paternostro}},\ }\bibfield  {title} {\bibinfo {title} {Irreversible entropy
  production, from quantum to classical},\ }\href
  {https://arxiv.org/abs/2009.07668} {\bibfield  {journal} {\bibinfo  {journal}
  {arXiv:2009.07668 [quant-ph]}\ } (\bibinfo {year} {2020})}\BibitemShut
  {NoStop}%
\bibitem [{\citenamefont {Kosloff}(2013)}]{2013_Entropy_15_02100_Kosloff}%
  \BibitemOpen
  \bibfield  {author} {\bibinfo {author} {\bibfnamefont {R.}~\bibnamefont
  {Kosloff}},\ }\bibfield  {title} {\bibinfo {title} {Quantum thermodynamics: a
  dynamical viewpoint},\ }\href {https://doi.org/10.3390/e15062100} {\bibfield
  {journal} {\bibinfo  {journal} {Entropy}\ }\textbf {\bibinfo {volume} {15}},\
  \bibinfo {pages} {2100} (\bibinfo {year} {2013})}\BibitemShut {NoStop}%
\bibitem [{\citenamefont {Tuncer}\ and\ \citenamefont
  {M\"{u}stecapl{\i}o\u{g}lu}(2020)}]{2020_AsliReview}%
  \BibitemOpen
  \bibfield  {author} {\bibinfo {author} {\bibfnamefont {A.}~\bibnamefont
  {Tuncer}}\ and\ \bibinfo {author} {\bibfnamefont {O.}~\bibnamefont
  {M\"{u}stecapl{\i}o\u{g}lu}},\ }\bibfield  {title} {\bibinfo {title} {Quantum
  thermodynamics and quantum coherence engines},\ }\href
  {https://doi.org/10.3906/fiz-2009-12} {\bibfield  {journal} {\bibinfo
  {journal} {Turk. J. Phys.}\ }\textbf {\bibinfo {volume} {44}},\ \bibinfo
  {pages} {404} (\bibinfo {year} {2020})}\BibitemShut {NoStop}%
\bibitem [{\citenamefont {Li}\ \emph {et~al.}(2012)\citenamefont {Li},
  \citenamefont {Ren}, \citenamefont {Wang}, \citenamefont {Zhang},
  \citenamefont {H\"anggi},\ and\ \citenamefont {Li}}]{2012_Phononics}%
  \BibitemOpen
  \bibfield  {author} {\bibinfo {author} {\bibfnamefont {N.}~\bibnamefont
  {Li}}, \bibinfo {author} {\bibfnamefont {J.}~\bibnamefont {Ren}}, \bibinfo
  {author} {\bibfnamefont {L.}~\bibnamefont {Wang}}, \bibinfo {author}
  {\bibfnamefont {G.}~\bibnamefont {Zhang}}, \bibinfo {author} {\bibfnamefont
  {P.}~\bibnamefont {H\"anggi}},\ and\ \bibinfo {author} {\bibfnamefont
  {B.}~\bibnamefont {Li}},\ }\bibfield  {title} {\bibinfo {title} {Colloquium:
  Phononics: Manipulating heat flow with electronic analogs and beyond},\
  }\href {https://doi.org/10.1103/RevModPhys.84.1045} {\bibfield  {journal}
  {\bibinfo  {journal} {Rev. Mod. Phys.}\ }\textbf {\bibinfo {volume} {84}},\
  \bibinfo {pages} {1045} (\bibinfo {year} {2012})}\BibitemShut {NoStop}%
\bibitem [{\citenamefont {Manzano}\ \emph {et~al.}(2020)\citenamefont
  {Manzano}, \citenamefont {Parrondo},\ and\ \citenamefont
  {Landi}}]{2020_arXiv_2011_04560}%
  \BibitemOpen
  \bibfield  {author} {\bibinfo {author} {\bibfnamefont {G.}~\bibnamefont
  {Manzano}}, \bibinfo {author} {\bibfnamefont {J.}~\bibnamefont {Parrondo}},\
  and\ \bibinfo {author} {\bibfnamefont {G.}~\bibnamefont {Landi}},\ }\bibfield
   {title} {\bibinfo {title} {Non-abelian quantum transport and thermosqueezing
  effects},\ }\href {https://arxiv.org/abs/2011.04560} {\bibfield  {journal}
  {\bibinfo  {journal} {arXiv:2011.04560 [quant-ph]}\ } (\bibinfo {year}
  {2020})}\BibitemShut {NoStop}%
\bibitem [{\citenamefont {Filipowicz}\ \emph {et~al.}(1986)\citenamefont
  {Filipowicz}, \citenamefont {Javanainen},\ and\ \citenamefont
  {Meystre}}]{Maser1986}%
  \BibitemOpen
  \bibfield  {author} {\bibinfo {author} {\bibfnamefont {P.}~\bibnamefont
  {Filipowicz}}, \bibinfo {author} {\bibfnamefont {J.}~\bibnamefont
  {Javanainen}},\ and\ \bibinfo {author} {\bibfnamefont {P.}~\bibnamefont
  {Meystre}},\ }\bibfield  {title} {\bibinfo {title} {Theory of a microscopic
  maser},\ }\href {https://doi.org/10.1103/PhysRevA.34.3077} {\bibfield
  {journal} {\bibinfo  {journal} {Phys. Rev. A}\ }\textbf {\bibinfo {volume}
  {34}},\ \bibinfo {pages} {3077} (\bibinfo {year} {1986})}\BibitemShut
  {NoStop}%
\bibitem [{\citenamefont {Scully}\ and\ \citenamefont
  {Zubairy}(1997)}]{ScullyZubairy1997}%
  \BibitemOpen
  \bibfield  {author} {\bibinfo {author} {\bibfnamefont {M.~O.}\ \bibnamefont
  {Scully}}\ and\ \bibinfo {author} {\bibfnamefont {M.~S.}\ \bibnamefont
  {Zubairy}},\ }\href {https://doi.org/10.1017/CBO9780511813993} {\emph
  {\bibinfo {title} {Quantum Optics}}}\ (\bibinfo  {publisher} {Cambridge
  University Press},\ \bibinfo {year} {1997})\BibitemShut {NoStop}%
\bibitem [{\citenamefont {Meystre}\ and\ \citenamefont
  {Sargent}(2007)}]{MeystreSargent2007}%
  \BibitemOpen
  \bibfield  {author} {\bibinfo {author} {\bibfnamefont {P.}~\bibnamefont
  {Meystre}}\ and\ \bibinfo {author} {\bibfnamefont {M.}~\bibnamefont
  {Sargent}},\ }\href {https://doi.org/10.1007/978-3-540-74211-1} {\emph
  {\bibinfo {title} {Elements of Quantum Optics}}},\ \bibinfo {edition} {4th}\
  ed.\ (\bibinfo  {publisher} {Springer-Verlag Berlin Heidelberg},\ \bibinfo
  {year} {2007})\BibitemShut {NoStop}%
\bibitem [{\citenamefont {Quan}\ \emph {et~al.}(2006)\citenamefont {Quan},
  \citenamefont {Zhang},\ and\ \citenamefont {Sun}}]{QThermalization2006}%
  \BibitemOpen
  \bibfield  {author} {\bibinfo {author} {\bibfnamefont {H.~T.}\ \bibnamefont
  {Quan}}, \bibinfo {author} {\bibfnamefont {P.}~\bibnamefont {Zhang}},\ and\
  \bibinfo {author} {\bibfnamefont {C.~P.}\ \bibnamefont {Sun}},\ }\bibfield
  {title} {\bibinfo {title} {Quantum-classical transition of photon-carnot
  engine induced by quantum decoherence},\ }\href
  {https://doi.org/10.1103/PhysRevE.73.036122} {\bibfield  {journal} {\bibinfo
  {journal} {Phys. Rev. E}\ }\textbf {\bibinfo {volume} {73}},\ \bibinfo
  {pages} {036122} (\bibinfo {year} {2006})}\BibitemShut {NoStop}%
\bibitem [{\citenamefont {Liao}\ \emph {et~al.}(2010)\citenamefont {Liao},
  \citenamefont {Dong},\ and\ \citenamefont {Sun}}]{QThermalization2010}%
  \BibitemOpen
  \bibfield  {author} {\bibinfo {author} {\bibfnamefont {J.-Q.}\ \bibnamefont
  {Liao}}, \bibinfo {author} {\bibfnamefont {H.}~\bibnamefont {Dong}},\ and\
  \bibinfo {author} {\bibfnamefont {C.~P.}\ \bibnamefont {Sun}},\ }\bibfield
  {title} {\bibinfo {title} {Single-particle machine for quantum
  thermalization},\ }\href {https://doi.org/10.1103/PhysRevA.81.052121}
  {\bibfield  {journal} {\bibinfo  {journal} {Phys. Rev. A}\ }\textbf {\bibinfo
  {volume} {81}},\ \bibinfo {pages} {052121} (\bibinfo {year}
  {2010})}\BibitemShut {NoStop}%
\bibitem [{\citenamefont {Li}\ \emph {et~al.}(2014)\citenamefont {Li},
  \citenamefont {Zou}, \citenamefont {Yu}, \citenamefont {Xu}, \citenamefont
  {Li},\ and\ \citenamefont {Shao}}]{QThermalization2014}%
  \BibitemOpen
  \bibfield  {author} {\bibinfo {author} {\bibfnamefont {H.}~\bibnamefont
  {Li}}, \bibinfo {author} {\bibfnamefont {J.}~\bibnamefont {Zou}}, \bibinfo
  {author} {\bibfnamefont {W.-L.}\ \bibnamefont {Yu}}, \bibinfo {author}
  {\bibfnamefont {B.-M.}\ \bibnamefont {Xu}}, \bibinfo {author} {\bibfnamefont
  {J.-G.}\ \bibnamefont {Li}},\ and\ \bibinfo {author} {\bibfnamefont
  {B.}~\bibnamefont {Shao}},\ }\bibfield  {title} {\bibinfo {title} {Quantum
  coherence rather than quantum correlations reflect the effects of a reservoir
  on a system's work capability},\ }\href
  {https://doi.org/10.1103/PhysRevE.89.052132} {\bibfield  {journal} {\bibinfo
  {journal} {Phys. Rev. E}\ }\textbf {\bibinfo {volume} {89}},\ \bibinfo
  {pages} {052132} (\bibinfo {year} {2014})}\BibitemShut {NoStop}%
\bibitem [{\citenamefont {Breuer}\ and\ \citenamefont
  {Petruccione}(2002)}]{BreuerAndPetruccione-2002}%
  \BibitemOpen
  \bibfield  {author} {\bibinfo {author} {\bibfnamefont {H.~P.}\ \bibnamefont
  {Breuer}}\ and\ \bibinfo {author} {\bibfnamefont {F.}~\bibnamefont
  {Petruccione}},\ }\href
  {https://doi.org/10.1093/acprof:oso/9780199213900.001.0001} {\emph {\bibinfo
  {title} {The theory of open quantum systems}}},\ \bibinfo {edition} {1st}\
  ed.\ (\bibinfo  {publisher} {Oxford University Press},\ \bibinfo {year}
  {2002})\ Chap.~\bibinfo {chapter} {3}, pp.\ \bibinfo {pages}
  {130--137}\BibitemShut {NoStop}%
\bibitem [{\citenamefont {Ghosh}\ \emph {et~al.}(2019)\citenamefont {Ghosh},
  \citenamefont {Mukherjee}, \citenamefont {Niedenzu},\ and\ \citenamefont
  {Kurizki}}]{2019_EPJSpecTop_Kurizki}%
  \BibitemOpen
  \bibfield  {author} {\bibinfo {author} {\bibfnamefont {A.}~\bibnamefont
  {Ghosh}}, \bibinfo {author} {\bibfnamefont {V.}~\bibnamefont {Mukherjee}},
  \bibinfo {author} {\bibfnamefont {W.}~\bibnamefont {Niedenzu}},\ and\
  \bibinfo {author} {\bibfnamefont {G.}~\bibnamefont {Kurizki}},\ }\bibfield
  {title} {\bibinfo {title} {Are quantum thermodynamic machines better than
  their classical counterparts?},\ }\href
  {https://doi.org/10.1140/epjst/e2019-800060-7} {\bibfield  {journal}
  {\bibinfo  {journal} {Eur. Phys. J. Spec. Top.}\ }\textbf {\bibinfo {volume}
  {227}},\ \bibinfo {pages} {2043} (\bibinfo {year} {2019})}\BibitemShut
  {NoStop}%
\bibitem [{\citenamefont {Carmichael}(1999)}]{Carmichael-1999}%
  \BibitemOpen
  \bibfield  {author} {\bibinfo {author} {\bibfnamefont {H.}~\bibnamefont
  {Carmichael}},\ }\href {\doibase 10.1007/978-3-662-03875-8} {\emph {\bibinfo
  {title} {Statistical Methods in Quantum Optics 1}}},\ \bibinfo {edition} {1st}\ ed.\ (\bibinfo
  {publisher} {Springer-Verlag},\ \bibinfo {address} {Berlin Heidelberg},\
  \bibinfo {year} {1999})\ Chap.~\bibinfo {chapter} {6}, pp.\ \bibinfo {pages}
  {193--209}\BibitemShut {NoStop}%
\end{thebibliography}
%

\end{document}